\documentclass[aps,prl,showpacs,preprintnumbers,amsmath,amssymb,groupedaddress]{revtex4}
\usepackage[dvips]{graphicx}
\usepackage{dcolumn}

\begin{document}

\title
{Emergent Antiferromagnetism in $D$-wave Superconductor with Strong Paramagnetic Pair-Breaking}

\author{Yuhki Hatakeyama and Ryusuke Ikeda}

\affiliation{%
Department of Physics, Kyoto University, Kyoto 606-8502, Japan
}

\date{\today}


\begin{abstract} 
It is theoretically shown that, in the four-fold symmetric $d$-wave superconducting phase, a paramagnetic pair-breaking (PPB) enhanced sufficiently by increasing the applied magnetic field induces not only the Fulde-Ferrell-Larkin-Ovchinnikov (FFLO) superconducting state but also an incommensurate antiferromagnetic (AFM) order with ${\bf Q}$-vector {\it parallel} to a gap node. This AFM ordering tends to occur only below $H_{c2}$ at low temperatures, i.e., in the presence of a nonvanishing superconducting energy gap $\Delta$ rather than in the normal phase. Through a detailed study on the resulting AFM order and its interplay with the FFLO spatial modulation of $\Delta$, it is argued that the strange high field and low temperature (HFLT) superconducting phase of CeCoIn$_5$ is a coexisting phase of the FFLO and incommensurate AFM orders, and that this PPB mechanism of an AFM ordering is also the origin of the AFM quantum critical fluctuation which has occurred close to $H_{c2}(0)$ in several unconventional superconductors 
including CeCoIn$_5$. 
\end{abstract}

\pacs{}


\maketitle

%

\section{I. Introduction}

Recently, the presence of an antiferromagnetic (AFM) quantum critical behavior near the superconducting (SC) depairing field (or, the mean field upper critical field) $H_{c2}(0)$ at low temperatures has been commonly found in several unconventional superconductors such as CeCoIn$_5$ \cite{QCP115Co1,QCP115Co2}, pressured CeRhIn$_5$ \cite{QCP115Rh}, NpPd$_2$Al$_5$ \cite{Haga}, Ce$_2$PdIn$_8$ \cite{Poland}, and Tl-compounds of cuprates \cite{Shibauchi}. Most of these materials belong to the so-called heavy-fermion superconductors and hence, are expected to have a large Zeeman term, i.e., remarkable Pauli paramagnetic pair breaking (PPB) effects. Conventionally, an AFM order is expected to be suppressed by the presence of a finite SC energy gap $|\Delta|$ \cite{Ueda,Machida} below the mean field SC phase transition, indicating a possibility of enhancement of AFM fluctuation or order {\it above} $H_{c2}(0)$. However, a closer examination of the AFM critical behavior suggests the presence of an AFM quantum critical point (QCP) {\it below} $H_{c2}(0)$ \cite{QCP115Co2}.  In fact, measurements in the SC state of the heavy fermion superconductor CeCoIn$_5$ showing a remarkably large PPB \cite{Bianchi2,Radovan,SIkeda} in finite magnetic fields ($H \neq 0$) clearly show the presence of AFM fluctuation in the SC state which is enhanced with {\it increasing} $H$ up to $H_{c2}$ \cite{Kasahara,Canada,Flouquet}. A schematic picture on the AFM critical fluctuation near $H_{c2}(0)$ is represented in Fig.1. 

The incommensurate AFM {\it order} discovered recently through neutron scattering measurements \cite{Kenzel1,Kenzel2} at the high field corner of the $H$-$T$ SC phase diagram of CeCoIn$_5$ in ${\bf H} \parallel ab$ will not be an independent event of the above-mentioned field-induced enhancement of incommensurate AFM fluctuation below $H_{c2}$. This AFM order has been detected just in the so-called high field low temperature (HFLT) phase of this material in ${\bf H} \perp c$ which has been previously identified \cite{Bianchi2,ada} with the long-sought spatially {\it modulated} Fulde-Ferrell-Larkin-Ovchinnikov (FFLO) SC state \cite{FFLO}. If focusing only on the magnetic properties seen through, e.g., the neutron scattering data \cite{Kenzel1,Kenzel2}, it might be natural to identify the HFLT phase with an incommensurate AFM phase coexisting with the spatially uniform $d$-wave SC order. However, the fact that the HFLT phase is destabilized by quite a small amount of not only magnetic impurities \cite{Tokiwa1} but also nonmagnetic ones \cite{Tokiwa2} is incompatible with the picture \cite{Kenzel1} favoring the presence of the uniform $d$-wave SC order in the HFLT phase. In fact, it has been found \cite{RI10} that such an unexpected impurity effect is consistent only with the picture identifying the HFLT phase with a SC state with one-dimensional modulation parallel to ${\bf H}$ \cite{ada}, such as a kind of the FFLO state. Then, it is necessary to clarify how this FFLO picture on the HFLT phase is compatible with the presence of the AFM order detected in the same phase. 
\begin{figure}[t]
\scalebox{0.6}[0.6]{\includegraphics{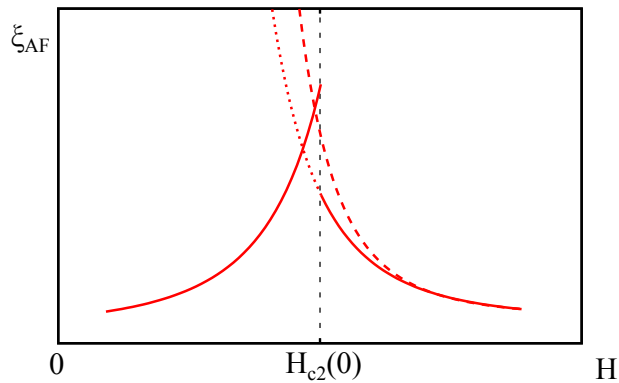}}
\caption{Field dependence of the AFM correlation length $\xi_{\rm AF}$ described schematically. For simplicity, the presence of a narrow HFLT phase just below $H_{c2}$ is neglected here (see sec.V and VI). The $\xi_{\rm AF}$ (solid) curve below $H_{c2}$ reflects the PPB-induced AFM ordering, while the solid curve above $H_{c2}$ results purely from the AFM fluctuation in the normal state. The dotted curve below $H_{c2}$ is the extrapolation of the high field curve to lower fields. If the $H_{c2}$-transition (the mean field SC transition) is of second order so that the SC fluctuation is not negligible above $H_{c2}$, the growth of $\xi_{\rm AF}$ upon decreasing the field in $H > H_{c2}$ should, as indicated by the dashed curve, become sharper as a consequence of the coupling between the two orderings (see eqs.(29) and (31) below). }
\end{figure}

In the present work, we develop a theory comprehensively explaining the above-mentioned phenomena occurring at the high field side of the $H$-$T$ SC phase diagram of a $d$-wave superconductor with strong PPB such as CeCoIn$_5$. To be specific, the favorable direction of the expected staggered moment is assumed throughout this paper to be perpendicular to the basal plane, i.e., parallel to the $c$-axis of the tetragonal structure of the quasi 2D SC materials according to an observation Ref.\cite{Kenzel1} on CeCoIn$_5$. It is found by extending the conventional model on coexistence of SC and AFM orders to the nonzero field ($H \neq 0$) case with strong PPB that, in $d$-wave superconductors with strong PPB, an incommensurate AFM order tends to be realized in higher fields but below $H_{c2}(0)$ if the AFM ${\bf Q}$-vector in the commensurate limit, called ${\bf Q}_0$ hereafter, is parallel to the nodal direction of the SC energy gap. Throughout this paper, the component of ${\bf Q}_0$ in the $a$-$b$ plane of a tetragonal crystal is ($\pi$, $\pi$) so that the SC pairing state with a gap node parallel to ${\bf Q}_0$ is inevitably the $d_{x^2-y^2}$-paired one. 

This PPB-induced AFM ordering or fluctuation has two crucial implications. First of all, the present AFM order occurs more easily in the SC phase with finite SC energy gap than in the normal state and thus, clarifies why no situation with AFM order only in the normal phase above $H_{c2}(0)$ is seen in those materials \cite{QCP115Co1,QCP115Co2,QCP115Rh,Haga,Poland,Shibauchi} with an AFM critical behavior around $H_{c2}(0)$. Second of all, since the presence or absence of the field-induced AFM ordering or fluctuation strongly depends on the relative orientation between the gap node direction and ${\bf Q}_0$, the present theory is also useful for addressing the pairing symmetry of those SC materials including CeCoIn$_5$ : In the case with ${\bf Q}_0$ parallel to ($\pi$, $\pi$), the presence of the field-induced AFM fluctuation near $H_{c2}(0)$ implies the $d_{x^2-y^2}$-pairing symmetry of the SC material. 

The present field-induced AFM ordering is a characteristic property of superconductors with strong PPB together with the FFLO SC state and the first order $H_{c2}$-transition \cite{Izawa,ada}, and hence, a consistent emergence of both this AFM order and the FFLO SC order in the same material is naturally expected. In fact, a couple of observations \cite{Tokiwa1,Tokiwa2,RI10,Watanabe,RI071,Miclea,RI072} in the HFLT phase of CeCoIn$_5$ have been done so far which support identification between the HFLT phase of this material and a FFLO state with a modulation parallel \cite{ada,RI10} to the applied field ${\bf H}$, which will be called a longitudinal FFLO state hereafter \cite{com0}. For this reason, we also examine the field-induced AFM ordering in the longitudinal FFLO state by invoking the Pauli limit, in which the orbital pair-breaking inducing the vortices is neglected, to evaluate thermodynamics and construct a $H$-$T$ phase diagram. It is found that the FFLO spatial modulation significantly extends the AFM ordered region so that the situation in which the AFM order is absent outside the FFLO state in the phase diagram is easily realized. The obtained results on the phase diagram will be compared with recent NMR data \cite{Kumagai11} presenting a firm evidence of {\it both} the longitudinal FFLO structure and the AFM order in the HFLT phase of CeCoIn$_5$ in ${\bf H} \perp c$. A preliminary report of the present work can be found in Refs.\cite{IHA} and \cite{sces}. We note that the presence of PPB-induced AFM fluctuation has been first noticed through our numerical studies of effects of an AFM quantum critical fluctuation on the vortex form factor \cite{IHA,aoyama}.  

When we discuss the phase diagram including a field-induced AFM ordering in the text, a situation with an indication of the presence of an AFM QCP $H^*$ will not be distinguished from the case with a finite AFM transition temperature in some field range. In fact, the field at which the AFM transition temperature is the highest in the latter situation would correspond to $H^*$ if the former situation is realized, and the former is easily realized from the latter, i.e., by reducing the strength of the electron repulsion leading to the AFM ordering in the normal state. 

This paper is organized as follows. In sec.II, two theoretical methods for studying the AFM ordering in the $d$-wave superconductors are explained in details. In sec.III and IV, numerical results on the AFM ordering following from them are presented, and their implications are explained by assuming the $d$-wave SC phase not to include a FFLO spatial modulation. The corresponding results in a FFLO phase are considered in sec.V. Section VI is devoted to a summary of the present work and final remarks. 

\section{II. Model} 

In the present study, we have examined a possible AFM ordering according to two approaches to be explained below. In zero magnetic field ($H=0$), our starting electronic model is the same in the two approaches and can be expressed by the Hamiltonian including just the two interaction channels of a $d$-wave superconductivity and antiferromagnetism 
${\cal H} = {\cal H}_{\mathrm{kin}} + {\cal H}_{\mathrm{AFM}} + {\cal H}_{\mathrm{SC}}$, where  
\begin{eqnarray}
{\cal H}_{\mathrm{kin}} &=& d \sum_{\sigma, j} \int \!d^2{\bf r}_\perp\ \biggl[ 
[\psi_j^{(\sigma)}({\bf r}_\perp) ]^\dagger \, \varepsilon_\perp(-i \nabla_\perp)  \psi_j^{(\sigma)} ({\bf r}_\perp) - \frac{J}{2} \biggl( [\psi_j^{(\sigma)}({\bf r}_\perp) ]^\dagger \psi_{j+1}^{(\sigma)}({\bf r}_\perp) + {\rm h.c.} \biggr) \biggr], \nonumber \\
{\cal H}_{\mathrm{AF}} &=& - U \sum_{{\bf q}, {\hat {\bf n}}} S_{\hat {\bf n}}^\dagger ({\bf q}) S_{\hat {\bf n}}({\bf q}), \nonumber \\
{\cal H}_{\mathrm{SC}} &=& -\frac{|g|}{4} \sum_{{\bf q}} \, \sum_{{\bf k},\sigma} w^*_{{\bf k}} c^\dagger_{{\bf k},\sigma} c^\dagger_{- {\bf k}+{\bf q},-\sigma} \sum_{{\bf k}^\prime,\sigma^\prime} w_{{\bf k}^\prime} c_{-{\bf k}^\prime+{\bf q},-\sigma^\prime} c_{{\bf k}^\prime,\sigma^\prime}
\label{originalH}
\end{eqnarray}
with 
\begin{eqnarray}
\psi_j^{(\sigma)}({\bf r}_\perp) &=& \frac{1}{\sqrt{N d L_x L_y}} \sum_{{\bf k}} c_{{\bf k},\sigma} e^{i({\bf k}_\perp\cdot{\bf r}_\perp + k_c j d)},  \nonumber \\
S_{{\hat {\bf n}}}({\bf q}) &=& \sum_{{\bf k}, \alpha, \beta} c^\dagger_{{\bf k}, \alpha} ({\hat \sigma}\cdot{\hat {\bf n}})_{\alpha, \beta} c_{{\bf k}+{\bf Q}_0+{\bf q}, \beta}. 
\label{psidef}
\end{eqnarray} 
In the above expressions, $\sigma$ ($= \pm 1$) denotes the spin projection, and the gap function satisfies the property $w_{{\bf k}+{\bf Q}_0} = -w_{{\bf k}}$ peculiar to the $d_{x^2-y^2}$-pairing state. Further, in the case of a Fermi surface with perfect nesting, the relation $\varepsilon({\bf k}+{\bf Q}_0) = - \varepsilon({\bf k})$ is satisfied. Through our explanation of our theoretical expressions, the unit $\hbar=c=k_B=1$ will be used. 

In the presence of a uniform magnetic field ${\bf H}$, the spin quantization axis parallel to ${\bf H}$ will be chosen hereafter. Further, the dispersion relation $\varepsilon(-i\nabla_\perp)$ needs to be replaced by 
$\varepsilon(-i \nabla_\perp + e {\bf A}) + I \sigma$, where $I$ is the Zeeman energy and is usually written as $\mu_{\rm B} g H$ with a $g$-factor and the Bohr magneton $\mu_{\rm B}$. 

For the moment, we focus on the mean field approximation neglecting both the SC and AFM fluctuations, and roles of the AFM {\it fluctuation} will be discussed in sec.IV and V. The AFM staggered field $m$ playing the role of the AFM, or spin density wave, order parameter is given by $m({\bf q})=U \langle S_{\hat {\bf n}}({\bf q}) \rangle$, where $\langle \,\,\, \rangle$ denotes the statistical average. 
When the AFM staggered moment carries a finite momentum ${\bf q}$, ${\bf Q} \equiv {\bf Q}_0+{\bf q}$ expresses the incommensurate AFM modulation wave vector. Expecting a finite ${\bf q}$ to be uniquely chosen at the microscopic level, the mean field expression of ${\cal H}_{\rm AF}$ may be represented by 
\begin{equation}
{\cal H}_{\rm {AF,MF}} = U^{-1} |m_{\hat {\bf n}}({\bf q})|^2 - [m_{\hat {\bf n}}(-{\bf q}) S_{\hat {\bf n}}({\bf q}) + {\rm h.c.}].
\label{MFAF}
\end{equation}
Here, the expression was written for general ${\hat {\bf n}}$. In the ensuing analysis, we assume ${\hat {\bf n}}$ to be fixed to the $c$-axis of the tetragonal structure for any direction of the magnetic field based on an experimental report on this issue \cite{Kenzel1} on CeCoIn$_5$. Thus, we will not write the index ${\hat {\bf n}}$ in $m_{\hat {\bf n}}$ hereafter. Implication of this assumption on the fixed ${\hat {\bf n}}$ will be commented on in sec.III. 
Similarly, the corresponding mean field expression of ${\cal H}_{\rm SC}$ is 
\begin{equation}
{\cal H}_{\rm {SC,MF}} = |g|^{-1} |\Delta(0)|^2 - \biggl[ \Delta^*(0) \sum_{{\bf k},\sigma} \frac{w_{\bf k}}{2} \sigma c_{-{\bf k}, -\sigma} \, c_{{\bf k}, \sigma} + {\rm h.c.} \biggr].
\label{MFSC}
\end{equation}
Here, the SC order parameter satisfying the gap equation 
\begin{equation}
\Delta = \frac{|g|}{2} \sum_{{\bf k},\sigma} w_{\bf k} \sigma \langle c_{-{\bf k},-\sigma} c_{{\bf k},\sigma} \rangle 
\label{gapeq}
\end{equation}
was, for convenience of description, assumed to be spatially uniform. 
We perform the mean field analysis of ${\cal H}$ in the following two 
approaches separately. 
One is a perturbative approach based on the microscopic derivation of a Ginzburg-Landau (GL) free energy of a form expanded in powers of {\it both} the SC and AFM order parameters. This method is useful in discussing, at least, the case with a relatively weaker PPB in which the $H_{c2}$-transition remains second order, because the orbital pair-breaking effect of the magnetic field inducing the vortices, necessary for obtaining the second order $H_{c2}$-transition, is included in the familiar manner. We note that the second order $H_{c2}$-transition is not a consequence of the GL expansion in the SC order parameter, as have been demonstrated through a derivation of the first order $H_{c2}$-transition \cite{ada}. However, this method is insufficient for examining the detailed structure of the AFM order reflecting the quasiparticle's dispersion relation. Another approach is to focus on the Pauli limit in which the field-induced vortices are completely neglected. The neglect of the vortices is inappropriate for considering response properties in the SC phase such as the vortex elasticity, while it may not affect evaluation of thermodynamic quantities of superconductors with strong PPB significantly. Rather, the details of the AFM order can be examined numerically within this approach. Note that these two approaches are complementary with each other. 

\subsection{Perturbative Approach}

First, let us start with explaining the perturbative approach. The quasiparticle energy incorporated in ${\cal H}_{\rm kin}$ is assumed in this approach to satisfy 
\begin{equation}
\varepsilon({\bf k}+{\bf Q}_0) = - \varepsilon({\bf k}) - T_{c0} 
\delta_{\rm IC}, 
\label{disp}
\end{equation} 
where the dimensionless parameter $\delta_{\rm IC}$ measures the incommensurability, i.e., the deviation from the perfect nesting condition for a spin-density-wave or AFM ordering, ${\bf Q}_0=(\pi/a, \pi/a, \pi/d)$ with the lattice constants $a$ (in the $ab$ direction) and $d$ (in the $c$-direction) is the commensurate AFM modulation wavevector. Note that, here, $\delta_{\rm IC}$ is assumed to be a constant. The general case in which $\delta_{\rm IC}$ is ${\bf k}$-dependent will be studied in the Pauli limit approach to be given later. 
Further, the opened Fermi surface in the $c$-direction, i.e., the quasi 2D nature, might become important in some electronic processes, 
while it is safely negligible in considering spatial variations of the order parameter fields $\Delta$ in a vortex state and of $m$ as far as the spatial anisotropy of the SC material to be defined later is relatively small. For this reason, the order parameter fields will be assumed \cite{RI071} to be functions of a continuous spatial coordinate ${\bf r}=({\bf r}_\perp, z)$ by replacing $jd$ by $z$. 

Regarding the Green's function ${\cal G}^{(\sigma)}(\tau; {\bf r}_\perp,{\bf r}_\perp^\prime; j, j') \equiv - \langle \mathrm{T}_\tau [ \psi^{(\sigma)}_j({\bf r}_\perp,\tau) (\psi^{(\sigma)}_{j'}({\bf r}_\perp^\prime,0))^\dagger] \rangle$ defined in the {\it normal} state, the quasi classical approximation for ${\cal G}^{(\sigma)}_{\varepsilon_n}({\bf r}_\perp, {\bf r}_\perp^\prime; j,j') = \int_0^\beta \!d\tau\ {\cal G}^{(\sigma)}(\tau; {\bf r}_\perp, {\bf r}_\perp^\prime) e^{i\varepsilon_n \tau}$, i.e., 
\begin{eqnarray}
{\cal G}^{(\sigma)}_{\varepsilon_n}({\bf r}_\perp, {\bf r}_\perp^\prime; j,j')
&\simeq& {\cal G}^{(\sigma)}_{\varepsilon_n}({\bf r}_\perp-{\bf r}_\perp^\prime; j-j')|_{H=0} \exp\biggl( ie \int^{{\bf r}_\perp}_{{\bf r}^\prime_\perp} {\bf A}({\bf s}) \cdot d{\bf s} \biggr), 
\label{semicl}
\end{eqnarray}
will be used, where $\varepsilon_n= \pi (2n+1)/\beta$ is a Fermion Matsubara frequency, and $\beta=1/T$ is the inverse temperature. In diagrammatic calculations, the formula 
\begin{equation}
\exp\left( -2ie \int^{{\bf r}_1}_{\bf r} {\bf A}({\bf s}) \cdot d{\bf s} \right) \Delta({\bf r}_1) = \exp\left( i({\bf r}_1-{\bf r}) \cdot {\bf \Pi} \right) \Delta({\bf r}) 
\label{Wherthamer}
\end{equation}
with ${\bf \Pi} = -i\nabla_\perp - 2e {\bf A}({\bf r})$ and 
\begin{eqnarray}
{\cal G}^{(\sigma)}_{\varepsilon_n}({\bf k}) &=& d \sum_j \int \!d^2 {\bf r}_\perp\ {\cal G}^{(\sigma)}_{\varepsilon_n}({\bf r}_\perp-{\bf r}_\perp^\prime; j-j')|_{H=0}  e^{-i {\bf k}_\perp \cdot ({\bf r}_\perp - {\bf r}_\perp^\prime)-i k_cd(j-j')} 
= \frac{1}{i\varepsilon_n - \varepsilon({\bf k}) + I\sigma}
\label{Green}
\end{eqnarray}
will be used. 
The orbital pair-breaking effect is incorporated through the relation (\ref{semicl}). 

The contributions associated with the SC and AFM orderings to the free energy density are given by the sum 
$f_{\mathrm{GL}} = f^{(2)}_{\Delta} + f^{(2)}_{m} + f^{(4)}_{\Delta} + f^{(4)}_m + f_{\Delta m}^{(2,2)} + f_{\Delta m}^{(2,4)}$. The last sixth order term $\propto |\Delta|^2 m^4$ is not negligible in evaluating the sign of $m^4$ term in the free energy in the SC state. The purely SC contributions take the form 
\begin{eqnarray}
f^{(2)}_\Delta &=& \biggl\langle \Delta^*({\bf r}) \biggl[ \frac{1}{|g|} - K^{(2)}_{\Delta}({\bf \Pi}) \biggr] \Delta({\bf r}) \biggr\rangle_{sp}, \nonumber \\
f^{(4)}_{\Delta} &=& \biggl\langle K^{(4)}_{\Delta}( {\bf \Pi}_i ) \Delta^*({\bf r}_1) \Delta({\bf r}_2) \Delta^*({\bf r}_3) \Delta({\bf r}_4) |_{{\bf r}_i \to {\bf r}} \biggr\rangle_{sp}, 
\label{FSC}
\end{eqnarray}
where $\langle \,\,\, \rangle_{sp}$ denotes the spatial average, and 
\begin{eqnarray}
K^{(2)}_{\Delta}({\bf \Pi}) &=& \frac{1}{2 \beta} \sum_{n,{\bf k},\sigma} |w_{{\bf k}}|^2 {\cal G}^{(\sigma)}_{\varepsilon_n}({\bf k}) \, {\cal G}^{(-\sigma)}_{-\varepsilon_n}(-{\bf k}+{\bf \Pi}), \nonumber \\
K^{(4)}_{\Delta}( {\bf \Pi}_i ) \!\! &=& \!\! \frac{1}{2 \beta} \sum_{n,{\bf k},\sigma} |w_{\bf k}|^4 {\cal G}^{(\sigma)}_{\varepsilon_n}({\bf k}) \, {\cal G}^{(-\sigma)}_{-\varepsilon_n}(-{\bf k}+{\bf \Pi}_2)  {\cal G}^{(\sigma)}_{\varepsilon_n}({\bf k}+{\bf \Pi}^*_3-{\bf \Pi}_2) \, {\cal G}^{(-\sigma)}_{-\varepsilon_n}(-{\bf k}+{\bf \Pi}^*_1). 
\label{KSC}
\end{eqnarray}
On the other hand, $f^{(2)}_m$ and $f^{(4)}_m$ are the corresponding GL terms in the AFM order parameter $m$ under vanishing $\Delta$, and, for simplicity, its $q=0$ case will be described here : 
\begin{eqnarray}
f_{m}^{(2)} &=& \left[ \frac{1}{U} - \frac{1}{2 \beta} \sum_{n,{\bf k},\sigma} {\cal G}^{(\sigma)}_{\varepsilon_n}({\bf k}) \, {\cal G}^{(\bar{\sigma})}_{\varepsilon_n}({\bf k}+{\bf Q}_0) \right] m^2, \nonumber \\
f_{m}^{(4)} &=& \biggl[ \frac{1}{2 \beta} \sum_{n, {\bf k},\sigma} {\cal G}^{(\sigma)}_{\varepsilon_n}({\bf k}) \, {\cal G}^{({\bar{\sigma}})}_{\varepsilon_n}({\bf k}+{\bf Q}_0) \, {\cal G}^{(\sigma)}_{\varepsilon_n}({\bf k})  {\cal G}^{(\bar{\sigma})}_{\varepsilon_n}({\bf k}+{\bf Q}_0) \biggr] m^4. 
\label{FAF}
\end{eqnarray}
Further, the coupling term between the two orders in the free energy density takes the form 
\begin{eqnarray}
f_{\Delta \, m}^{(2,2)} &=& \biggl\langle \biggl[ 2 K_{\Delta \, m, \, 1}( {\bf \Pi}_i ) + K_{\Delta \, m, \, 2}( {\bf \Pi}_i ) \biggr] \Delta^*({\bf r}_1)  \Delta({\bf r}_2) m^2 |_{{\bf r}_i \to {\bf r}} \biggr\rangle_{sp} 
\label{FAFSC}
\end{eqnarray}
where 
\begin{eqnarray}
K_{\Delta \, m, \, 1}( {\bf \Pi}_i ) &=& - \beta^{-1} \sum_{n,{\bf k}, \sigma} |w_{\bf k}|^2 {\cal G}^{(\sigma)}_{\varepsilon_n}({\bf k}+{\bf Q}_0) \, {\cal G}^{({\bar{\sigma}})}_{\varepsilon_n}({\bf k})  {\cal G}^{(-\bar{\sigma})}_{-\varepsilon_n} (-{\bf k} + {\bf \Pi}_2) \, {\cal G}^{\bar{\sigma}}_{\varepsilon_n}({\bf k}-{\bf \Pi}_2+{\bf \Pi}^*_1), \nonumber \\
K_{\Delta \, m, \, 2}( {\bf \Pi}_i ) &=& - \beta^{-1} \sum_{n, {\bf k}, \sigma} w_{\bf k} w_{{\bf k}+{\bf Q}_0}^* {\cal G}^{(\sigma)}_{\varepsilon_n}({\bf k}+{\bf \Pi}_2) \, {\cal G}^{(-\sigma)}_{-\varepsilon_n}(- {\bf k}) {\cal G}^{(-\bar{\sigma})}_{-\varepsilon_n}(-{\bf k}+{\bf Q}_0) \, {\cal G}^{(\bar{\sigma})}_{\varepsilon_n}({\bf k}+{\bf Q}_0 
+ {\bf \Pi}^*_1). 
\label{KAFSC}
\end{eqnarray}
In the above expressions, $\bar{\sigma}$ is $\sigma$ in ${\hat {\bf n}} \parallel {\bf H}$ and $-\sigma$ in ${\hat {\bf n}} \perp {\bf H}$, respectively. 

At this stage, one of key observations in the present work can be explained by noting that ${\cal G}^{(-\sigma)}_{-\varepsilon_n}({\bf k}+{\bf Q}_0) = - {\cal G}^{(\sigma)}_{\varepsilon_n}({\bf k})$ in the commensurate ($\delta_{\rm IC} \to 0$) limit:  In ${\hat {\bf n}} \parallel {\bf H}$ case, it is noticed through comparison with eq.(\ref{KSC}) that $K_{\Delta \, m, \, n}$ ($n=1$ and $2$) become the same expression as $K^{(4)}_{\Delta}$, except the presence of ${\bf \Pi}$s' operations, and thus that, as well as the sign change of $K^{(4)}_{\Delta}$ in higher fields resulting in the first order $H_{c2}$-transition \cite{ada,MT}, they also become negative as PPB is enhanced. Since it implies that the $m^2$-term in the free energy is reduced with increasing PPB, coexistence of the SC and AFM orders is favored, or an AFM QCP becomes closer with increasing $H$ or by an enhancement of PPB. This conclusion is not limited to the $d_{x^2-y^2}$-pairing case and is satisfied for any pairing state. In contrast, the corresponding mechanism of a PPB-induced AFM ordering in the SC state in ${\hat {\bf n}} \perp {\bf H}$ is peculiar to the $d_{x^2-y^2}$-pairing case and will be 
explained later. 

To perform the ${\bf k}$-integrals, a particle-hole symmetry will be assumed to be approximately satisfied around the Fermi surface by introducing a constant density of states $N(0)$ on the Fermi surface as a useful parameter. By replacing ${\varepsilon}({\bf k}+{\bf \Pi})$ by $\varepsilon({\bf k}) + {\bf v}_{\bf k}\cdot{\bf \Pi}$ with the velocity ${\bf v}_{\bf k}$ on the Fermi 
surface, we obtain 
\begin{eqnarray}
K^{(2)}_{\Delta}({\bf \Pi}) &=& \frac{N(0)}{2 \beta} \sum_{n,\sigma} \int \!d\varepsilon({\bf k})\ \langle |w_{\bf k}|^2 {\cal G}_{\varepsilon_n}^{(\sigma)}({\bf k}) \, {\cal G}_{-\varepsilon_n}^{(-\sigma)}(-{\bf k}) \rangle_{\mathrm{FS}} 
= \pi \beta^{-1} N(0) \sum_{n, \sigma} \biggl\langle \frac{i\mathop{\mathrm{sgn}}(\varepsilon_n) |w_{\bf k}|^2}{2i\varepsilon_n+2I\sigma - {\bf v}_{\bf k}\cdot{\bf \Pi}} \biggr\rangle_{\mathrm{FS}} \nonumber \\
&=& 2 \pi t N(0) \int_0^\infty \!d\rho\ f(\rho) \biggl\langle |w_{\bf k}|^2 \exp\biggl(-i \rho \frac{{\bf v}_{\bf k}\cdot{\bf \Pi}}{T_{c0}} \biggr) 
\biggr\rangle_{\mathrm{FS}} 
\label{Kernel2}
\end{eqnarray}
where $\langle \,\,\, \rangle_{\rm FS}$ denotes the angle average over the Fermi surface, 
$f(\rho)= {\rm cos}(2 I \rho/T_{c0})/{\rm sinh}(2 \pi t \rho)$, $t=1/(\beta T_{c0})$, ${\bf v}({\bf k}) = \partial \varepsilon({\bf k})/\partial {\bf k}$, and the parameter integral 
\begin{equation}
\frac{1}{\kappa} = \int_0^\infty \!d\rho\ e^{-\kappa\rho} \quad (\mathop{\mathrm{Re}}\kappa > 0) 
\end{equation} 
was used. 
As usual, the zero field SC transition temperature $T_{c0}$ is defined by 
\begin{equation} 
\frac{1}{N(0)|g|}={\rm ln}\biggl(\frac{1}{\beta T_{c0}} \biggr)+\sum_{\varepsilon_n > 0}^{\varepsilon_c} \frac{2 \pi}{\beta \varepsilon_n}, 
\label{eq:cutoff}
\end{equation}
where $\varepsilon_c$ is a high energy cut-off. 

In the presence of the orbital pair-breaking, the SC gap is varying in real space due to the presence of vortices. As far as the PPB is not extremely strong in the ballistic limit \cite{RI072}, the vortex lattice solution may be described by the lowest Landau level mode of $\Delta$. Then, we have 
\begin{equation}
\Delta({\bf r}) = \Delta \varphi_0({\bf r}). 
\end{equation}
Here, $\varphi_0$ is the familiar Abrikosov state \cite{Eilenbergerfl} 
\begin{equation}
\varphi_0({\bf r}) = \biggl( \frac{k^2}{\pi} \biggr)^{\frac{1}{4}} \sum_{s=-\infty}^\infty \exp\biggl[ i \biggl( \frac{sk}{r_H} y + \frac{\pi}{2} s^2 \biggr) - \frac{1}{2} \biggl( \frac{x}{r_H} + sk \biggr)^2 \biggr] 
\end{equation}
expressed in terms of the integer $s$, which satisfies the normalization condition $\langle |\varphi_0|^2 \rangle_{sp}=1$. Further, $r_H = 1/\sqrt{2|eH|}$ is the magnetic length for the Cooper pairs, 
and $k = \pi^{1/2}/3^{1/4}$ for the triangular lattice. 
Noting that $\Pi_\pm = r_H (\Pi_x \pm i\Pi_y)/\sqrt{2}$ is the raising and lowering operator for the Landau levels satisfying $[\Pi_-, \Pi_+]=1$, we have 
$\exp(i T_{c0}^{-1} {\bf {v_k}}\cdot{\bf {\Pi}} \rho) = \exp(- |\eta|^2/2) e^{i \eta \Pi_+} e^{i \eta^* \Pi_-}$, where 
$\eta = \rho(v_x-iv_y)/(\sqrt{2} r_H T_{c0})$. 
Using the property 
$\langle \varphi^*_0({\bf r}) e^{-i \rho T_{c0}^{-1} {\bf v}_{\bf k}\cdot{\bf \Pi}} \varphi_0({\bf r}) \rangle_{sp} = \exp(- |\eta|^2/2)$, the quadratic term in the free energy is expressed in terms of eq.(\ref{Kernel2}) by 
\begin{eqnarray}
f_{\Delta}^{(2)} &=& N(0) \biggl[ {\rm ln}\biggl(\frac{1}{T_{c0} \beta} \biggr) + 2 \pi t \int_0^\infty \!d\rho\  \biggl\langle |w_{{\bf k}}|^2 \biggl( \frac{1}{{\rm sinh}(2 \pi t \rho)} - f(\rho) \exp\biggl(- \frac{{\bf v}_\perp^2}{4 r_H^2 T_{c0}^2} \rho^2 \biggr) \biggr) \biggr\rangle_{\mathrm{FS}} \biggr] \langle |\Delta|^2 \rangle_{sp} 
\label{F2SC}
\end{eqnarray}
where ${\bf v}_\perp^2 = v_x^2+v_y^2$ in ${\bf H} \parallel c$. 

Deriving the corresponding expression of the SC quartic term $F_\Delta^{(4)}$ is performed in a similar manner. The kernel in $F_\Delta^{(4)}$ takes the form 
\begin{widetext}
\begin{eqnarray}
K^{(4)}_{\Delta}( {\bf \Pi}_i ) &=& \pi T N(0) \sum_{n,\sigma} \biggl\langle \frac{-i{\rm sgn}(\varepsilon_n) |w_{{\bf k}}|^4}
{[2 i \varepsilon_n + 2I \sigma + {\bf v}_{\bf k}\cdot{\bf \Pi}^*_1] [2 i \varepsilon_n + 2 I \sigma + {\bf v}_{\bf k}\cdot{\bf \Pi}_2] [2 i \varepsilon_n + 2 I \sigma + {\bf v}_{\bf k}\cdot{\bf \Pi}^*_3]} \biggr\rangle_{FS} 
+ ({\bf \Pi}_2 \leftrightarrow {\bf \Pi}_4) \nonumber \\
&=& \frac{2 \pi t N(0)}{T_{c0}^2} \int_0^\infty \!\prod_{i=1}^3 d\rho_i\ f(\rho_1+\rho_2+\rho_3) \langle |w_{\bf k}|^4 \, e^{i T_{c0}^{-1} (\rho_1 {\bf v}_{\bf k}\cdot{\bf \Pi}^*_1+\rho_2{\bf v}_{\bf k}\cdot{\bf \Pi}_2+\rho_3{\bf v}_{\bf k}\cdot{\bf \Pi}^*_3)} \rangle_{\mathrm{FS}} 
+ ({\bf \Pi}_2 \leftrightarrow {\bf \Pi}_4). 
\end{eqnarray}
By using the identity 
\begin{eqnarray}
\exp\biggl(i\rho \frac{{\bf v}_{\bf k}\cdot{\bf \Pi}}{T_{c0}} \biggr) \, \varphi_0 = \biggl( \frac{k^2}{\pi} \biggr)^{\frac{1}{4}} \sum_{s=-\infty}^\infty \exp\biggl[-\frac{1}{2} (|\eta|^2 - \eta^2) 
+ i \biggl( \frac{s k}{r_H} y + \frac{\pi}{2} s^2 \biggr) - \frac{1}{2} \biggl(\frac{x}{r_H} + sk +\sqrt{2} \eta \biggr)^2 \biggr], 
\end{eqnarray}
we have 
\begin{eqnarray}
f^{(4)}_{\Delta} &=& 2 \pi t N(0) T_{c0}^2 \biggl[ \int_0^\infty \! \prod_{i=1}^3 d\rho_i\ f(\rho_1+\rho_2+\rho_3) \frac{k}{\sqrt{2\pi}} \sum_{l_1,l_2} (-1)^{l_1l_2} \exp\biggl[-\frac{1}{2} (l_1^2+l_2^2) k^2 \biggr] \nonumber \\ 
&\times& \biggl\langle |w_{\bf k}|^4 \exp\biggl[-\frac{1}{2} (|\eta_1|^2+|\eta_2|^2+|\eta_3|^2) \biggr] \mathop{\mathrm{Re}}[e^{-p_0}] \biggr\rangle_{\mathrm{FS}} \biggr] \biggl\langle \biggl( \frac{|\Delta|}{T_{c0}} \biggr)^4 \biggr\rangle_{sp}, 
\label{F4SC}
\end{eqnarray}
where 
\begin{equation}
p_0 = \frac{1}{2} ({\eta^*_1}^2+{\eta_2}^2+{\eta^*_3}^2) - \frac{1}{4} (\eta_2-\eta^*_1-\eta^*_3)^2 - \frac{k}{\sqrt{2}} [ l_1 (\eta_2-\eta^*_1+\eta^*_3) + l_2 (\eta_2+\eta^*_1-\eta^*_3) ], 
\end{equation}
$\eta_i = \rho_i (v_x-iv_y)/(\sqrt{2}r_H T_{c0})$ in ${\bf H} \parallel c$, 
and 
${\bf \Pi}_4 = {\bf \Pi^*}_1-{\bf \Pi}_2+{\bf \Pi}^*_3$. 

In ${\bf H} \perp c$, the corresponding expressions to eqs.(\ref{F2SC}) and (\ref{F4SC}) are given by replacing $v_x$ and $v_y$ in ${\bf H} \parallel c$ by $\gamma^{-1/2} v_y$ and $\gamma^{1/2} v_z$, respectively, where $\gamma = (\langle v_y^2 \rangle_{\mathrm{FS}}/\langle v_z^2 \rangle_{\mathrm{FS}})^{1/2} = 2 E_{\rm F} (1 - J/E_{\rm F})^{1/2}/(\pi J)$ is the anisotropy of the SC length scales, and $E_{\rm F}$ is the Fermi energy in 2D ($J \to 0$) limit. 

Next, the quadratic term $f_m^{(2)}$ will be rewritten by assuming the AFM staggered field to be uniform. A sign change of $f_m^{(2)}$ determines a position of the second order transition to the AFM phase if the O($m^4$) term in the free energy density is positive (see below). 
By defining the Neel temperature $T_{\rm N}$ in the normal phase according to eq.(\ref{eq:cutoff}) with $|g|$ and $T_{c0}$ replaced by $U$ and $T_{\rm N}$, respectively, 
$f_{m}^{(2)}$ in ${\hat {\bf n}} \parallel c \parallel {\bf H}$ becomes 
\begin{equation} 
f_{m}^{(2)} = N(0) \biggl[ {\rm ln}\biggl( \frac{1}{\beta T_{\rm N}} \biggr)+\frac{1}{2} \sum_{\sigma} {\rm Re} \biggl[ \psi\biggl( \frac{1}{2}+i\frac{I \sigma \beta}{2 \pi} + i \frac{\delta_{\rm IC} \beta}{4 \pi} \biggr) - \psi \biggl( \frac{1}{2} \biggr) \biggr] \biggr] 
 m^2
\label{eq:m2para}
\end{equation}
where $\psi(z)$ is the digamma function 
\begin{equation} 
\psi(z) = - \gamma + \sum_{n=0}^\infty \biggl( \frac{1}{n+1} -\frac{1}{n+z} \biggr). 
\end{equation}
The Zeeman energy term plays a similar role to the incommensurability $\delta_{\rm IC}$ for the AFM ordering in this field configuration. 

On the other hand, in ${\hat {\bf n}} \perp {\bf H}$, 
$f_{m}^{(2)}$ becomes 
\begin{equation} 
f_{m}^{(2)} = N(0) \biggl[ {\rm ln}\biggl( \frac{1}{\beta T_{\rm N}} \biggr) + {\rm Re} \biggl[ \psi\biggl( \frac{1}{2}+i\frac{\delta_{\rm IC} \beta}{4 \pi} \biggr) - \psi \biggl( \frac{1}{2} \biggr) \biggr] \biggr] 
m^2 
\label{eq:m2perp}.
\end{equation} 
Within the present model in which no ${\bf k}$-dependent anisotropy is assumed in the Zeeman energy term, the AFM transition temperature is $H$-independent even in the normal phase in this field configuration.

Now, let us turn to evaluating the kernels in the coupling term $f_{\Delta \, m}^{(2,2)}$. To be specific, in this subsection, we focus on the $d_{x^2-y^2}$-pairing case in which $w_{\bf k}=-w_{{\bf k}+{\bf Q}_0}$. The corresponding results in the $d_{xy}$-pairing 
case are given by changing the overall sign of $K_{\Delta \, m, \, 2}$ in the following expressions. 
In the ${\hat {\bf n}} \parallel H \parallel c$ case, we have 
\begin{eqnarray}
K_{\Delta \, m, \, 1}( {\bf \Pi}_i ) &=& -2 \pi \beta^{-1} N(0) \sum_{n,\sigma} \biggl\langle \frac{i\mathop{\mathrm{sgn}}(\varepsilon_n) |w_{\bf k}|^2}{[2i \varepsilon_n+2I\sigma+{\bf v}_{\bf k}\cdot{\bf \Pi^*}_1][2 i \varepsilon_n+2I\sigma+{\bf v}_{\bf k}\cdot{\bf \Pi}_2][2 i \varepsilon_n+2I\sigma+ T_{c0} \delta_{\rm IC} ]} \nonumber \\
&+& \frac{i\mathop{\mathrm{sgn}}(\varepsilon_n) |w_{\bf k}|^2}{[2 i \varepsilon_n+2I\sigma+{\bf v}_{\bf k}\cdot{\bf \Pi}^*_1] [2i\varepsilon_n+2I\sigma + T_{c0} \delta_{\rm IC}][2i\varepsilon_n+2I\sigma + T_{c0} \delta_{\rm IC} + {\bf v}_{\bf k}\cdot({\bf \Pi}^*_1-{\bf \Pi}_2)]} 
\biggr\rangle_{\mathrm{FS}} \nonumber \\ 
&=& 2 \pi t N(0) T_{c0}^{-2} \int_0^\infty \!\prod_{i=1}^3 d\rho_i\ f(\sum_{i=1}^3 \rho_i) \biggl\langle |w_{\bf k}|^2 \biggl[ e^{i\delta_{\rm IC} \rho_3} e^{i T_{c0}^{-1} {\bf v}_{\bf k}\cdot(\rho_1 {\bf \Pi}^*_1+\rho_2 {\bf \Pi}_2)} \nonumber \\
&+& e^{i \delta_{\rm IC} (\rho_2+\rho_3)} e^{i T_{c0}^{-1} {\bf v}_{\bf k}\cdot((\rho_1+\rho_3) {\bf \Pi}^*_1-\rho_3 {\bf \Pi}_2)} \biggr] + \mathrm{h.c.} \biggr\rangle_{\mathrm{FS}}, \nonumber \\ 
K_{\Delta \, m, \, 2}( {\bf \Pi}_i ) &=& - 2 \pi t N(0) T_{c0}^{-2} \int_0^\infty \!\prod_{i=1}^3 d\rho_i\ f(\sum_{i=1}^3 \rho_i) \biggl\langle |w_{\bf k}|^2 \biggl[ e^{i \delta_{\rm IC} (-\rho_2+\rho_3)} e^{i T_{c0}^{-1} {\bf v}_{\bf k}\cdot((\rho_1+\rho_3){\bf \Pi}^*_1-\rho_3 {\bf \Pi}_2)} \nonumber \\ 
&+& e^{i \delta_{\rm IC} (\rho_2-\rho_3)} e^{i T_{c0}^{-1} {\bf v}_{\bf k}\cdot(\rho_2 {\bf \Pi}^*_1-(\rho_1+\rho_2){\bf \Pi}_2)} \biggr] + \mathrm{h.c.} \biggr\rangle_{\mathrm{FS}}
\label{KSCAFM}
\end{eqnarray}
\end{widetext}

Then, the coupling term of the free energy is expressed, 
using the relation 
$\langle e^{i T_{c0}^{-1} {\bf v}_{\bf k}\cdot(\rho_1 {\bf \Pi}^*_1+\rho_2{\bf \Pi}_2)} \varphi^*_0({\bf r}_1) \varphi_0({\bf r}_2) |_{{\bf r}_i\to{\bf r}} \rangle_{sp} = e^{-(1/2)(|\eta_1|^2+|\eta_2|^2+2\eta^*_1\eta_2)}$, in the form 
\begin{widetext}
\begin{eqnarray}
f_{\Delta \, m}^{(2,2)} &=& 2 \pi t N(0) T_{c0}^2 \biggl[ \int_0^\infty \!\prod_{i=1}^3 d\rho_i\ f(\rho_1+\rho_2+\rho_3) \biggl\langle |w_{\bf k}|^2 [ 4 {\rm cos}(\delta_{\rm IC} \rho_3) e^{-(\rho_1+\rho_2)^2 ((v_\perp)^2/(2r_H T_{c0})^2} \nonumber \\
&-& 8 \sin(\delta_{\rm IC} \rho_2) \sin(\delta_{\rm IC} \rho_3) e^{-\rho_1^2 (v_\perp)^2/(2r_H T_{c0})^2} ] \biggr\rangle_{\mathrm{FS}} \biggr] \biggl\langle \frac{|\Delta|^2}{T_{c0}^2} \frac{m^2}{T_{c0}^2} \biggr\rangle_{sp}
\label{FSCAFM} 
\end{eqnarray}

On the other hand, in the ${\bf H} \perp c$ and ${\hat {\bf n}} \parallel c$ case, the corresponding expressions to eqs.(\ref{KSCAFM}) and (\ref{FSCAFM}) are 
\begin{eqnarray}
K_{\Delta \, m, \, 1}( {\bf \Pi}_i) 
	&=& \frac{N(0)}{T_{c0}^2} \int_0^\infty \!\prod_{i=1}^3 d\rho_i\ \frac{2\pi t}{\sinh[2\pi t(\rho_1+\rho_2+\rho_3)]} \biggl\langle |w_{\bf k}|^2 \biggl[ {\rm cos}\biggl(2 \frac{I}{T_{c0}} (\rho_1+\rho_2) \biggr) e^{i \delta_{\rm IC} \rho_3}e^{i T_{c0}^{-1} {\bf v}_{\bf k}\cdot(\rho_1 {\bf \Pi}^*_1+\rho_2 {\bf \Pi}_2)} \nonumber \\ 
&+& {\rm cos}\biggl(2 \frac{I}{T_{c0}} \rho_2 \biggr) e^{i \delta_{\rm IC} (\rho_1+ \rho_3)} e^{i T_{c0}^{-1} {\bf v}_{\bf k}\cdot((\rho_1+\rho_2){\bf \Pi}^*_1-\rho_1 {\bf \Pi}_2)} \biggr] + \mathrm{h.c.} \biggr\rangle_{\mathrm{FS}}, \nonumber \\ 
K_{\Delta \, m, \, 2}( {\bf {\Pi}}_i) &=&
	- \frac{N(0)}{T_{c0}^2} \int_0^\infty \!\prod_{i=1}^3 d\rho_i\ \frac{2\pi t}{\sinh[2 \pi t(\rho_1+\rho_2+\rho_3)]} \biggl\langle |w_{\bf k}|^2 \biggl
[ {\rm cos}\biggl(2 \frac{I}{T_{c0}}(\rho_1-\rho_2) \biggr) e^{-i \delta_{\rm IC} \rho_3} e^{i T_{c0}^{-1} {\bf v}_{\bf k}\cdot(\rho_1 {\bf \Pi}^*_1-\rho_2 {\bf \Pi}_2)} \nonumber \\ 
&+& {\rm cos}\biggl(2 \frac{I}{T_{c0}}(\rho_1-\rho_2) \biggr) e^{i \delta_{\rm IC} \rho_3} e^{i T_{c0}^{-1} {\bf v}_{\bf k}\cdot((\rho_1+\rho_3) {\bf \Pi}^*_1-(\rho_2+\rho_3){\bf \Pi}_2)} \biggr] 
	+ \mathrm{h.c.} \biggr\rangle_{\mathrm{FS}},
\end{eqnarray}
and 
\begin{eqnarray}
f_{\Delta \, m}^{(2,2)} &=& 
	N(0) T_{c0}^2 \biggl[ \int_0^\infty \!\prod_{i=1}^3 d\rho_i\ \frac{2\pi t}{\sinh[2\pi t(\rho_1+\rho_2+\rho_3)]} \biggl\langle |w_{\bf k}|^2 
\biggl[ 4 \cos\biggl(2I \frac{\rho_1+\rho_2}{T_{c0}} \biggr) \cos(\delta_{\rm IC} \rho_3) e^{- (\rho_1+\rho_2)^2 v_\perp^2/(2r_H T_{c0})^2} \nonumber \\
&+& 4 \cos\biggl(2 I \frac{\rho_2}{T_{c0}} \biggr) \cos(\delta_{\rm IC} (\rho_1+\rho_3)) e^{- {\rho_2}^2 v_\perp^2/(2 r_H T_{c0})^2} \nonumber \\
&-& 4 \cos\biggl(2 I \frac{\rho_1-\rho_2}{T_{c0}} \biggr) \cos(\delta_{\rm IC} \rho_3) e^{-(\rho_1-\rho_2)^2 (v_\perp)^2/(2 r_H T_{c0})^2} ] \biggr\rangle_{\mathrm{FS}} \biggr] \biggl\langle \frac{|\Delta|^2}{T_{c0}^2} \biggr\rangle_{sp}
\frac{m^2}{T_{c0}^2}. 
\end{eqnarray}
\end{widetext}

The above $K_{\Delta \, m, \, n}$ ($n=1$ and $2$) expressions clarify how the AFM ordering in the SC state in ${\hat {\bf n}} \perp {\bf H}$ case occurs: For simplicity, let us consider again the commensurate case with vanishing $\delta_{\rm IC}$ and neglect the orbital pair-breaking. Then, it is easily found that, in low $T$ limit, $K_{\Delta \, m, \, 1}$ approaches the positive value $N(0)/(2 I^2)$, while $-K_{\Delta \, m, \, 2}$ shows the logarithmic divergence $I^{-2} N(0)  {\rm ln}(\pi^{-1} I/T)$. Note that the term appearing through the anomalous Green's functions grows with a minus sign. Since this overall minus sign of $K_{\Delta \, m, \, 2}$ inducing an AFM ordering is a consequence of the property $w_{{\bf k}+{\bf Q}_0} = - w_{\bf k}$ in the $d_{x^2-y^2}$ pairing case, this tendency of an AFM ordering in the SC state is peculiar to a $d$-wave pairing symmetry with a gap node parallel to the expected ${\bf Q}_0$-vector of a commensurate AFM order. Of course, the divergence indicated above in low $T$ limit is, strictly speaking, an artifact of the use of the GL expansion with respect to $\Delta$. In section IV, however, it will be shown that the corresponding growth of the anomalous term of the coupled term of AFM and SC orders with a negative sign is satisfied beyond the GL expansion and thus that the PPB-induced AFM ordering in ${\hat {\bf n}} \perp {\bf H}$ should occur generally in any $d_{x^2-y^2}$-wave paired superconductor with strong PPB under a high magnetic field. 

To examine the character of the AFM transition, the terms quartic in $m$ have to be examined. In this perturbative approach, they consist of the normal contribution $f_m^{(4)}$ and the additional term $f_{\Delta m}^{(4,2)}$ including the SC contribution of O($|\Delta|^2$). As a broad tendency, the SC contribution $f_{\Delta m}^{(4,2)}$ seems to make the AFM transition a continuous one even if, as is seen in ${\bf H} \parallel c$ in some cases, $f_m^{(4)}$ is negative. The expressions of these two terms in the free energy density 
will be given in Appendix.

\subsection{Pauli limit}

In the charged systems leading to superconductivity at low temperatures, the two field-induced pair-breaking processes, the spin effect (i.e., PPB) and the orbital one inducing the vortices, need to be taken into account. In particular, when studying fluctuation effects and SC response properties such as the elastic responses peculiar to the vortex states, as pointed out elsewhere \cite{RIprl04}, the use of the Pauli limit in which the orbital pair-breaking effect is neglected could lead to erroneous results on physical properties. On the other hand, most of thermodynamic properties and the high field phase diagram of bulk superconductors with strong PPB are, in the mean field approximation, reasonably described by the Pauli limit \cite{ada,MT}. Paying attention to this point, we present here a formulation in the Pauli limit of the FFLO states and the AFM ordering in superconductors with strong PPB where the $H_{c2}$-transition is expected to be of first order at low temperatures. For simplicity, we assume here a 2D Fermi surface for which $J=0$. 

The Matsubara Green's functions will be defined in the form 
\begin{eqnarray}
G^{(\sigma)}(\tau; {\bf r}_\perp,{\bf r}_\perp^\prime) &=& - \langle T_\tau [ \psi^{(\sigma)}({\bf r}, \tau) [\psi^{(\sigma)}]^\dagger({\bf r}',0) ] 
\rangle, \nonumber \\ 
{\overline F^{(\sigma)}}(\tau; {\bf r}_\perp,{\bf r}_\perp^\prime) &=& - \langle T_\tau [ ([\psi^{(-\sigma)}]^\dagger({\bf r}, \tau) [\psi^{(\sigma)}]^\dagger({\bf r}',0) ] \rangle, \nonumber \\
F^{(\sigma)}(\tau; {\bf r}_\perp,{\bf r}_\perp^\prime) &=& - \langle T_\tau [ \psi^{(\sigma)}({\bf r}, \tau) \psi^{(-\sigma)}({\bf r}',0) ] \rangle, \nonumber \\
{\overline G}^{(\sigma)}(\tau; {\bf r}_\perp,{\bf r}_\perp^\prime) &=& - \langle T_\tau[ [\psi^{(\sigma)}]^\dagger({\bf r},\tau) \psi^{(\sigma)}({\bf r}',0) ] \rangle, 
\end{eqnarray}
where the notation of the Green's functions has been changed to avoid a unnecessary confusion. Alternatively, they will be used often in the matrix form 
\begin{eqnarray}
{\hat G}^{(\sigma)} = \biggl[ 
\begin{array}{cc} 
G^{(\sigma)} & F^{(\sigma)} \\ {\overline F}^{(\sigma)} & {\overline G}^{(-\sigma)} \\ 
\end{array} \biggr].
\end{eqnarray}
Then, the Fourier-transform of ${\hat G}^{(\sigma)}$, ${\hat G}^{(\sigma)}_{\varepsilon_n}({\bf k}; {\bf R}) \equiv \int d\tau e^{i\varepsilon_n \tau} \int d^3({\bf r}_\perp - {\bf r}_\perp^\prime) {\hat G}^{(\sigma)}(\tau; {\bf r}, {\bf r}^\prime) e^{-i{\bf k}\cdot({\bf r}_\perp - {\bf r}_\perp^\prime)}$, where ${\bf R}= ({\bf r}+{\bf r}^\prime)/2$, satisfies 
\begin{widetext}
\begin{eqnarray}
\biggl[ \begin{array}{cc}
        i\varepsilon_n - \varepsilon({\bf k})+I\sigma & -\Delta_{\bf k}({\bf R})\sigma \\
        \Delta_{\bf k}^*({\bf R})\sigma & -i\varepsilon_n-\varepsilon({\bf k})-I\sigma \\
        \end{array} \biggr] {\hat G}^{(\sigma)}_{\varepsilon_n}({\bf k},{\bf R}) 
= {\hat 1} + [{\bf v}_{\bf k}\cdot\partial_{\bf R}] {\hat G}^{(\sigma)}_{\varepsilon_n}({\bf k},{\bf R})
\end{eqnarray}
\end{widetext}
where $\Delta_{\bf k}({\bf R}) = \Delta({\bf R}) w_{\bf k}$, and the derivative operator will be defined as 
\begin{eqnarray}
\partial_{\bf R} = \biggl\{ \begin{array}{cc}
        {\bf \Pi}=-i \nabla_{\bf R} - 2e {\bf A}({\bf R}) & \mbox{for }\Delta({\bf R}) \\ 
        {\bf \Pi}^\dagger = - i \nabla_{\bf R} + 2e {\bf A}({\bf R}) & \mbox{for }\Delta^*({\bf R}) \\ 
        -i\nabla_{\bf R} & \mbox{otherwise}
        \end{array} 
\end{eqnarray}

Next, since a possibility of the FFLO state is considered, the Green's function will be expanded in powers of $\partial_{\bf R}$ in the way 
\begin{equation}
{\hat G}^{(\sigma)} = 
{\hat G}^{(\sigma)}_{(0)}
 + {\hat G}^{(\sigma)}_{(2)} + {\hat G}^{(\sigma)}_{(4)} 
+ \cdots,
\label{Gexp}
\end{equation}
where ${\hat G}^{(\sigma)}_{(n)}$ is the $n$-th order term of ${\hat G}$ in the gradient. The terms with odd $n$ have been neglected above which do not contribute to the free energy density. 
Each term in the expansion (\ref{Gexp}) is given by 
\begin{eqnarray}
{\hat G}^{(\sigma)}_{\varepsilon_n, \, (0)}({\bf k},{\bf R}) 
&=& \biggl[ \begin{array}{cc}
        i\varepsilon_n-\varepsilon({\bf k}) + I \sigma & -\Delta_{\bf k}({\bf R})\sigma \\
        \Delta_{{\bf k}}^*({\bf R})\sigma & -i\varepsilon_n-\varepsilon({\bf k})-I\sigma \\
        \end{array} \biggr]^{-1} = \frac{1}{D} \biggl[ \begin{array}{cc}
        -i\varepsilon_n-\varepsilon({\bf k})-I\sigma & \Delta_{\bf k}({\bf R})\sigma \\
        -\Delta_{\bf k}^*({\bf R})\sigma & i\varepsilon_n-\varepsilon({\bf k})+I\sigma \\
        \end{array} \biggr],
\end{eqnarray}
where 
\begin{equation}
D \equiv [\varepsilon({\bf k})]^2-(i\varepsilon_n+I\sigma)^2+|\Delta_{\bf k}|^2,
\end{equation}
and 
\begin{eqnarray}
{\hat G}^{(\sigma)}_{\varepsilon_n, \, (2)}({\bf k},{\bf R}) &=& {\hat G}^{(\sigma)}_{(0)} \biggl( {\bf v}_{\bf k}\cdot\partial_{\bf R} \biggl(
	{\hat G}^{(\sigma)}_{(0)} {\bf v}_{\bf k}\cdot\partial_{\bf R} {\hat G}^{(\sigma)}_{(0)} \biggr) \biggr) \nonumber \\
{\hat G}^{(\sigma)}_{\varepsilon_n, \, (4)}({\bf k},{\bf R}) &=& {\hat G}^{(\sigma)}_{(0)} \biggl( {\bf v}_{\bf k}\cdot\partial_{\bf R} \biggl(
{\hat G}^{(\sigma)}_{(0)} {\bf v}_{\bf k}\cdot\partial_{\bf R} \biggl( {\hat G}^{(\sigma)}_{(0)} {\bf v}_{\bf k}\cdot\partial_{\bf R} \biggl(
{\hat G}^{(\sigma)}_{(0)} {\bf v}_{\bf k}\cdot\partial_{\bf R} {\hat G}^{(\sigma)}_{(0)} \biggr) \biggr) \biggr) \biggr).  
\end{eqnarray}

In writing down the expression of the free energy, the magnitude of the SC energy gap $|\Delta|$ has been assumed to be much more rigid compared with that of a possible AFM order parameter. This approximation is reasonable when the $H_{c2}$-transition is discontinuous. 
Then, $|\Delta|$ may be determined selfconsistently just from the SC part of the free energy density $f_{\Delta}$, which becomes \cite{EilenbergerGor}
\begin{eqnarray} 
f_{\Delta} &=& \biggl\langle  \frac{|\Delta({\bf R})|^2}{|g|} + \frac{1}{2 \beta} \sum_{\varepsilon_n, {\bf k},\sigma} \int^{\infty \mathop{\mathrm{sgn}}(\varepsilon_n)}_{\varepsilon_n} \!d\omega\  \mathop{\mathrm{Tr}} \biggl[ i {\hat \sigma}_z {\hat G}^{(\sigma)}_{\omega} ({\bf k},{\bf R}) \biggr] \biggr\rangle_{sp}. 
\end{eqnarray}
As well as ${\hat G}$, $f_{\Delta}$ may also be classified in the form of gradient expansion \cite{EilenbergerGor}
\begin{widetext}
\begin{eqnarray}
f_{\Delta}^{(0)} &=& \biggl\langle \frac{|\Delta|^2}{|g|} - \beta^{-1} \sum_{\varepsilon_n>0} \sum_{\bf k}
	\ln \biggl[\frac{(\varepsilon_n^2+[\varepsilon({\bf k})]^2+|\Delta_{\bf k}|^2-I^2)^2+4 \varepsilon_n^2 I^2}{(\varepsilon_n^2+[\varepsilon({\bf k})]^2-I^2)^2+4 \varepsilon_n^2 I^2} \biggr] \biggr\rangle_{sp}, \nonumber \\ 
f_{\Delta}^{(2)} &=& \beta^{-1} \biggl\langle \sum_{\varepsilon_n>0} \sum_{\bf k} \biggl[\frac{a_s^2-b_s^2}{(a_s^2+b_s^2)^2} |{\bf v}_{\bf k}\cdot{\bf \Pi} \Delta_{\bf k}|^2 \nonumber \\
&+& \frac{2}{3} \frac{(2[\varepsilon({\bf k})]^2-\varepsilon_n^2+I^2-|\Delta_{\bf k}|^2) (a_s^4-6a_s^2b_s^2+b_s^4) - 4 a_s b_s^2 (a_s^2-b_s^2) }{(a_s^2+b_s^2)^4} ({\bf v}_{\bf k}\cdot\nabla |\Delta_{\bf k}|^2)^2 \biggr] \biggr\rangle_{sp}, 
\nonumber \\
f_{\Delta}^{(4)} &\simeq& \biggl\langle \beta^{-1} \sum_{\varepsilon_n>0} \sum_{\bf k} \biggl[ \frac{2}{3} \frac{(2 [\varepsilon({\bf k})]^2-\varepsilon_n^2+I^2-|\Delta_{\bf k}|^2) (a_s^4-6a_s^2b_s^2+b_s^4) - 4 a_s b_s^2 (a_s^2 - b_s^2)}{(a_s^2+b_s^2)^4}|({\bf v}_{\bf k}\cdot{\bf \Pi})^2 \Delta_{\bf k}|^2 \biggr] \biggr\rangle_{sp},
\label{FSCgrad}
\end{eqnarray}
\end{widetext}
where 
$a_s = [\varepsilon({\bf k})]^2+\varepsilon_n^2+|\Delta_{\bf k}|^2-I^2$, and $b_s = 2 \varepsilon_n I$. 

In our analysis in the Pauli limit, the only ${\bf R}$-dependence of the SC order parameter we consider is that of the longitudinal FFLO state 
\begin{equation}
\Delta({\bf r}) = \biggl[ \sqrt{2}\cos(q_{\rm LO} \, x) \biggr] \Delta
\label{LOdelta}
\end{equation}
in ${\bf H} \parallel {\hat x}$, where just a single Fourier component with the wave length $2 \pi/q_{\rm LO}$ is assumed for the FFLO modulation, while the two kinds of $d$-wave pairing symmetries 
\begin{eqnarray}
w_{\bf k} &=& \biggl\{ \begin{array}{cc} \cos(k_xa)-\cos(k_ya) & (d_{x^2-y^2} \mbox{-wave}) \\ \sin(k_xa) \sin(k_ya) & (d_{xy} \mbox{-wave}) \end{array}
\end{eqnarray}
will be considered. Further, to examine the details of the resulting AFM order in the SC state, the dispersion relation 
\begin{eqnarray}
\varepsilon({\bf k}) &=& - 2 t_1(\cos(k_xa)+\cos(k_ya)) - 4 t_2\cos(k_xa)\cos(k_ya) - 2 t_3(\cos(2k_xa)+\cos(2k_ya)) - \mu
\label{dispersionYamada}
\end{eqnarray}
will be used following Ref.\cite{HIkeda}. An incommensurability of AFM order primarily stems from a nonvanishing $t_2$ term. 

A possible AFM order may be considered in the form of a Landau expansion in the staggered field $m$ of the free energy density, in particular, if the AFM transition is of second order. Then, the AFM contributions in the free energy density take the form $f_m = f_m^{(2)} + f_m^{(4)}$ as a power series in $m$, where 
\begin{equation}
f_m^{(2)} = \biggl\langle \biggl[ U^{-1} - \chi^{(n)} - \chi^{(an)} \biggr] m^2 \biggr\rangle_{sp}. 
\label{fm2pauli}
\end{equation}
In ${\bf H} \perp {\hat {\bf n}}$ with ${\hat {\bf n}} \perp ab$, 
\begin{eqnarray}
\chi^{(n)} &=& - \beta^{-1} \sum_{n, {\bf k}, \sigma} G^{(\sigma)}_{\varepsilon_n, \, (0)}({\bf k}) \, G^{(-\sigma)}_{\varepsilon_n, \, (0)}({\bf k}+{\bf Q}_0), \nonumber \\
\chi^{(an)} &=& \beta^{-1} \sum_{n, {\bf k}, \sigma} F^{(\sigma)}_{\varepsilon_n, \, (0)}({\bf k}) \, {\overline F}^{(-\sigma)}_{\varepsilon_n, \, (0)}({\bf k}+{\bf Q}_0), 
\end{eqnarray}
and 
\begin{eqnarray}
f_m^{(4)} &=& \frac{1}{2 \beta} \sum_{\varepsilon_n, {\bf k}, \sigma} \frac{1}{2}{\rm Tr} \biggl[ {\hat \sigma}_z {\hat G}^{(\sigma)}_{\varepsilon_n, \, (0)}({\bf k}+{\bf Q}_0) {\hat \sigma}_z {\hat G}^{(\sigma)}_{\varepsilon_n, \, (0)}({\bf k})  {\hat \sigma}_z {\hat G}^{(\sigma)}_{\varepsilon_n, \, {(0)}}({\bf k} + {\bf Q}_0) {\hat \sigma}_z {\hat G}^{(\sigma)}_{\varepsilon_n, \, {(0)}}({\bf k}) \biggr] m^4. 
\end{eqnarray}
It is found that these expressions are rewritten in the form 
\begin{eqnarray}
\chi^{(n)} &=& \beta^{-1} \sum_{\varepsilon_n>0} \sum_{\bf k} \frac{4 [\varepsilon_n^2+I^2-\varepsilon({\bf k}) \varepsilon({\bf k}+{\bf Q}_0) ] a_\perp}{a_\perp^2+b_\perp^2}, \nonumber \\
\chi^{(an)} &=& - \beta^{-1} \sum_{\varepsilon_n>0} \sum_{\bf k} \frac{4 \Delta_{\bf k} \Delta^*_{{\bf k}+{\bf Q}_0} a_\perp}{a_\perp^2+b_\perp^2},
\end{eqnarray}
\begin{eqnarray}
a_\perp &=& (\varepsilon_n^2+[\varepsilon({\bf k})]^2+|\Delta_{\bf k}|^2-I^2) (\varepsilon_n^2+[\varepsilon({\bf k}+{\bf Q}_0)]^2 + |\Delta_{{\bf k}+{\bf Q}_0}|^2-I^2) + 4 \varepsilon_n^2 I^2, \nonumber \\ 
b_\perp &=& 2 \varepsilon_n I ([\varepsilon({\bf k}+{\bf Q}_0)]^2-[\varepsilon({\bf k})]^2+|\Delta_{{\bf k}+{\bf Q}_0}|^2-|\Delta_{\bf k}|^2), 
\end{eqnarray}

\begin{eqnarray}
f_{m}^{(2)} &=& \biggl[ \frac{1}{U} 
- \beta^{-1} \sum_{\varepsilon_n > 0} \sum_{\bf k} \frac{4 a_\perp}{a_\perp^2+b_\perp^2} (\varepsilon_n^2 + I^2 - \varepsilon({\bf k}) \varepsilon({\bf k}+{\bf Q}_0) - \Delta_{\bf k} \Delta^*_{{\bf k}+{\bf Q}_0} ) \biggr] m^2, 
\label{fm2perpn}
\end{eqnarray}

and 
\begin{widetext}
\begin{eqnarray}
f_m^{(4)} &=& 2 \beta^{-1} \sum_{\varepsilon_n>0} \sum_{\bf k} \frac{1}{(a_\perp^2+b_\perp^2)^2} \biggl\{ (a_\perp^2-b_\perp^2) \biggl[ (\varepsilon_n^2+I^2-\varepsilon({\bf k}) \, \varepsilon({\bf k}+{\bf Q}_0) - \Delta^*_{{\bf k}+{\bf Q}_0} \Delta_{\bf k} )^2 
\nonumber \\
&-& \varepsilon_n^2 [\varepsilon({\bf k}) - \varepsilon({{\bf k}+{\bf Q}_0})]^2 + I^2|\Delta_{\bf k}-\Delta_{{\bf k}+{\bf Q}_0}|^2 - |\varepsilon({\bf k}+{\bf Q}_0) \Delta_{\bf k}-\varepsilon({\bf k}) \Delta_{{\bf k}+{\bf Q}_0}|^2 \biggr] - 2a_\perp b_\perp^2 \biggr\} m^4. 
\label{fm4perpn}
\end{eqnarray}
\end{widetext}

In ${\hat {\bf n}} \parallel H \parallel c$, 
\begin{eqnarray}
\chi^{(n)} &=& \beta^{-1} \sum_{\varepsilon_n>0} \sum_{\bf k} \frac{4[ (\varepsilon_n^2-I^2-\varepsilon({\bf k}) \varepsilon({\bf k}+{\bf Q}_0) a_\parallel + 2 \varepsilon_n I b_\parallel]}{a_\parallel^2+b_\parallel^2} \nonumber \\
\chi^{(an)} &=& - \beta^{-1} \sum_{\varepsilon_n>0} \sum_{\bf k} \frac{4 \Delta_{\bf k} \Delta^*_{{\bf k}+{\bf Q}_0} a_\parallel}{a_\parallel^2 + 
b_\parallel^2}, 
\end{eqnarray}
where 
\begin{eqnarray}
a_\parallel &=& (\varepsilon_n^2+[\varepsilon({\bf k})]^2+|\Delta_{\bf k}|^2-I^2) (\varepsilon_n^2+[\varepsilon({\bf k}+{\bf Q}_0)]^2 + |\Delta_{{\bf k}+{\bf Q}_0}|^2-I^2) - 4\varepsilon_n^2 I^2 \nonumber \\
b_\parallel &=& 2 \varepsilon_n I (2 \varepsilon_n^2 - 2 I^2 + [\varepsilon({\bf k}+{\bf Q}_0)]^2 + [\varepsilon({\bf k})]^2 + |\Delta_{\bf k}|^2+|\Delta_{{\bf k}+{\bf Q}_0}|^2) 
\end{eqnarray}

\begin{widetext}
\begin{eqnarray}
f_{m}^{(2)} &=&  \biggl[ \frac{1}{U} - \beta^{-1} \sum_{\varepsilon_n>0} \sum_{\bf k} \frac{4 [(\varepsilon_n^2-I^2-[\varepsilon({\bf k})][\varepsilon({\bf k}+{\bf Q}_0)] -\Delta_{\bf k} \Delta^*_{{\bf k}+{\bf Q}_0}) a_\parallel + 2\varepsilon_n I b_\parallel]}{a_\parallel^2+b_\parallel^2} \biggr] m^2 
\end{eqnarray}

\begin{eqnarray}
f_m^{(4)} &=& \sum_{\varepsilon_n>0} \sum_{\bf k} \frac{2 \beta^{-1}}{(a_\parallel^2+b_\parallel^2)^2} \biggl\{ (a_\parallel^2-b_\parallel^2) \biggl[ (\varepsilon_n^2-I^2-\varepsilon({\bf k}) \varepsilon({\bf k}+{\bf Q}_0) + \Delta^*_{{\bf k}+{\bf Q}_0}\Delta_{\bf k})^2 - (\varepsilon_n^2-I^2) ([\varepsilon({\bf k}) + \varepsilon({\bf k}+{\bf Q}_0) ]^2 \nonumber \\
&+& |\Delta_{\bf k}-\Delta_{{\bf k}+{\bf Q}_0}|^2) 
- |\varepsilon({\bf k}+{\bf Q}_0) \Delta_{\bf k}+\varepsilon({\bf k}) \Delta_{{\bf k}+{\bf Q}_0}|^2 \biggr] - 4 \varepsilon_n I a_\parallel b_\parallel \biggl[  [ \varepsilon({\bf k}) + \varepsilon({\bf k}+{\bf Q}_0) ]^2+|\Delta_{\bf k}-\Delta_{{\bf k}+{\bf Q}_0}|^2 \nonumber \\
&-& 2(\varepsilon_n^2 - I^2 - \varepsilon({\bf k}) \varepsilon({\bf k}+{\bf Q}_0) + \Delta^*_{{\bf k}+{\bf Q}_0} \Delta_{\bf k}) \biggr] \biggr\} 
m^4
\end{eqnarray}
\end{widetext}

The expressions given in this section are used to examine the resulting $H$-$T$ phase diagram and the details of the expected AFM ordering. 

\begin{figure}[t]
\scalebox{0.6}[0.6]{\includegraphics{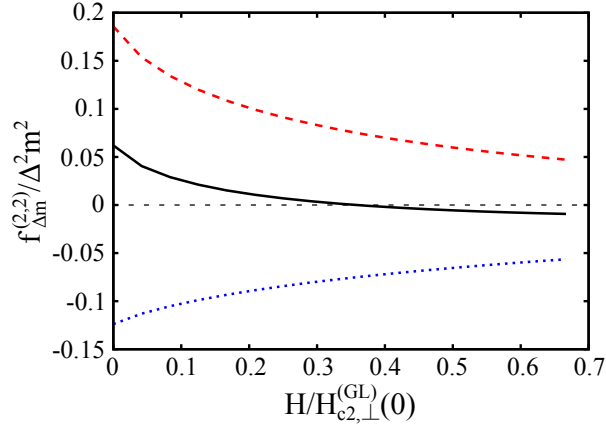}}
\caption{Field dependence of the coupling term $f_{\Delta \, m}^{(2,2)}$ (solid black curve) of the free energy density in ${\bf H} \perp c$ calculated in the perturbative approach and taken at $t=0.3$. The used parameters are $\alpha_{{\rm P}, \perp}=0.3$, $\delta_{\rm IC}=1.1$, and $\gamma=4.5$. The upper (red) dashed curve is the contribution from $K_{\Delta \, m, \, 1}$, while the lower (blue) dotted one is that from $K_{\Delta \, m, \, 2}$.}
\end{figure}

\section{III. Case with Second Order $H_{c2}$-Transition}

In this section, the $H$-$T$ phase diagram near $H_{c2}(0)$ and possible AFM ordering in the case with a moderately strong PPB will be numerically examined in terms of the theoretical expressions in the first half of the last section. In this case with a moderately strong PPB, the situation with first order $H_{c2}$-transition occurs rarely, and the $H_{c2}$-transition remains of second order even in low $T$ limit in most cases. This situation will be appropriate for explaining phenomena in the pressured CeRhIn$_5$ \cite{QCP115Rh}, Ce$_2$PdIn$_8$ \cite{Poland}, and other $d_{x^2-y^2}$-paired 
superconductors \cite{Shibauchi}. 

The strength of PPB is measured by the dimensionless parameter $\alpha_{\rm P}= I[H=H_{c2}^{({\rm GL})}(T=0)]/(2 \pi T_{c0})$, where $I(H)$ is the Zeeman energy. The so-called Maki parameter $\alpha_M$ corresponds to $\alpha_{\rm P}$ multiplied by the factor $7.1$. Since we focus here on the family of quasi 2D materials, the following two PPB 
parameters will be defined here in the manner depending on the direction of ${\bf H}$ 
\begin{eqnarray}
\alpha_{{\rm P}, \parallel} &=& \frac{I[H=H_{{c2},\parallel}^{({\rm GL})}(0)]}{2 \pi T_{c0}}, \nonumber \\
\alpha_{{\rm P}, \perp} &=& \frac{I[H=H_{{c2},\perp}^{({\rm GL})}(0)]}{2 \pi 
T_{c0}}, 
\end{eqnarray}
where $H_{{c2},\parallel}^{({\rm GL})}$ and $H_{{c2},\perp}^{({\rm GL})}= \gamma H_{{c2},\parallel}^{({\rm GL})}$ are the orbital depairing field in ${\bf H} \parallel c$ and ${\bf H} \perp c$, respectively. 

In this section, we show only calculation results obtained in terms of material parameters leading to an AFM order at finite temperatures. In systems with moderately strong PPB, no true AFM order has been detected so far, and just the presence of an AFM quantum critical fluctuation enhanced close to $H_{c2}(0)$ have been found. But, the field at which the AFM transition temperature is the highest would be transmuted to an apparent AFM QCP through a slight tuning of material parameters or including an introduction of impurity disorder. 

\begin{figure}[t]
\scalebox{0.7}[0.7]{\includegraphics{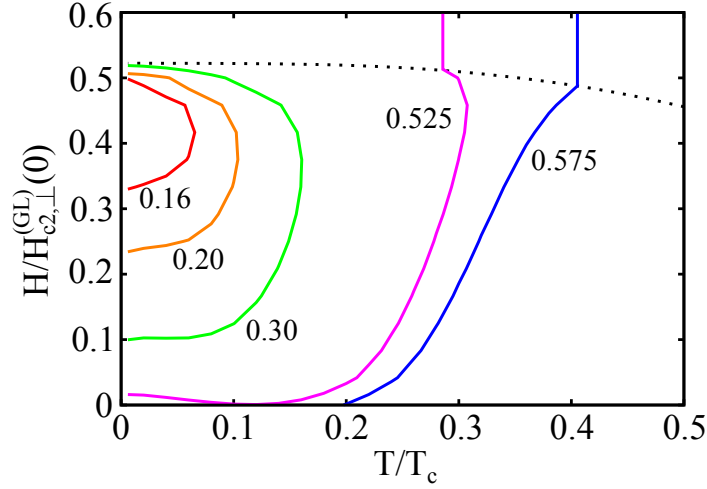}}
\caption{AFM transition curves described in $H$-$T$ phase diagram in ${\bf H} \perp c$ as a function of $T_{\rm N}$ which is the Neel temperature in the 
normal phase in the {\it commensurate} limit. Here, calculation was performed consistently with that in Fig.2. The AFM transition curves follow from the $T_{\rm N}/T_{c0}$ values in the range between $0.575$ (rightmost) and $0.16$ (leftmost). The AFM order in the normal state (i.e., in $H > H_{c2}(0)$) is absent when $T_{\rm N} < 0.32 T_{c0}$. The $H_{c2}$-transition on the dotted curve is of second order even at low $T$ because the orbital pair-breaking at this moderate value of $\alpha_{{\rm P}, \perp}$ is not negligible even in low $T$ limit.}
\end{figure}

\subsection{${\bf H} \perp c$}

First, calculated results in the in-plane field configuration ${\bf H} \perp c$  will be explained. A tendency of the PPB-induced AFM ordering is reflected in a field-induced sign change of the coupling or mixing term $f_{\Delta \, m}^{(2,2)}$ of the free energy density. An example of this sign change of $f_{\Delta \, m}^{(2,2)}(H)$ is given in Fig.2. In the figure, the field dependence to be seen in a conventional type II superconductor with a negligibly weak PPB is limited in low enough fields, $H/H_{\rm P} \leq 0.2$, where the field-induced reduction of $f_{\Delta \, m}^{(2,2)}$ will not be able to be distinguished from the conventional picture \cite{Ueda,Machida} on the AFM ordering stemming from a reduction of $f_{\Delta \, m}^{(2,2)}$ due to the field-induced reduction of $|\Delta|$. Namely, if the orbital depairing field lies in such a field range that the PPB is negligible for the disappearance of superconductivity, the field-induced AFM ordering would be regarded as being due to the vanishing of $|\Delta|$. As is seen below, however, a close inspection of the behaviors near $H_{c2}$ indicates that this PPB-induced AFM ordering is weakened by the field-induced reduction 
of $|\Delta|$. 

In Fig.3, AFM transition curves in the field range around $H_{c2}(0)$, obtained as a function of the Neel temperature $T_{\rm N}$ in the normal state in $H > H_{c2}$, are shown. Here, the same set of material parameters as in Fig.2 have been used, and the pressure dependence in real systems has been assumed to be directly reflected in that of $T_{\rm N}$ in the normal state. We expect Fig.3 to be comparable with the corresponding experimental phase diagrams on CeRhIn$_5$ (see Ref.\cite{QCP115Rh} and Figs.4 and 20 (a) of Ref.\cite{Knebel}). In the present electronic model, it is expected that not only the inverse of the electronic repulsive interaction, $1/U$, but also the incommensurability $|\delta_{\rm IC}|$ increase with increasing pressure. As mentioned previously \cite{IHA}, an increase of $|\delta_{\rm IC}|$ tends to enhance the PPB-induced AFM ordering (see also Fig.5 in sec.IV). In the case of pressured CeRhIn$_5$ \cite{QCP115Rh,Knebel}, however, it will be natural to expect that the pressure dependence of the AFM phase boundary is determined by that of $T_{\rm N}$ in the normal state \cite{com1}, while that of the incommensurability is a correction. Hereafter, we focus on the behavior in the SC state of Fig.3 that the AFM order disappears close to but below $H_{c2}(0)$. A couple of remarkable features are seen in Fig.3. First of all, at lower pressures (i.e., higher $T_{\rm N}$), the AFM transition temperature monotonously decreases with decreasing field. This behavior indicates that the reduction of the SC energy gap $|\Delta|$ rather than the PPB-origin plays a dominant role for an increase of the AFM transition point, because the orbital pair-breaking is more dominant than PPB at such high temperatures. As $T_{\rm N}$ is sufficiently lowered, however, the PPB primarily determines the field dependence of the AFM transition temperature, and the AFM phase appears just below $H_{c2}$ rather than above $H_{c2}$. Note that, very close to $H_{c2}$, the AFM order is rather lost in the present case where $H_{c2}$-transition is of second order. It implies that this AFM order is purely of SC origin and hence that it is {\it lost} as a consequence of the decrease of $|\Delta|$ on approaching $H_{c2}$ from below. This feature is one of the features clarifying that this AFM order is induced not by the decrease of $|\Delta|$ but rather by the PPB which is effective only in the SC state. 
\begin{figure}[t]
\scalebox{0.6}[0.6]{\includegraphics{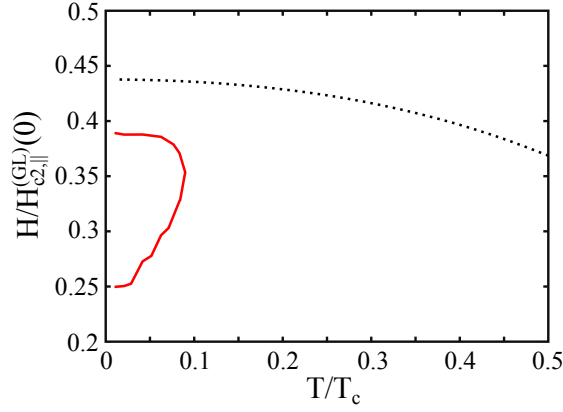}}
\caption{AFM transition curve described in $H$-$T$ phase diagram in ${\bf H} \parallel c$ and obtained in terms of $\gamma=4.5$, $\alpha_{{\rm P}, \parallel}=0.39$, $\delta_{\rm IC}=1.1$, and $T_{\rm N}/T_{c0}=0.6$. The $H_{c2}$-transition on the dotted curve is of second order. }
\end{figure}

\subsection{${\bf H} \parallel c$} 

Next, the corresponding results on the PPB-induced AFM ordering in ${\bf H} \parallel {\hat {\bf n}}$ and ${\hat {\bf n}} \parallel c$ 
will be briefly discussed. 
In contrast to the case in ${\bf H} \perp {\hat {\bf n}}$, the AFM ordering in the normal state in this case is suppressed, as well as the SC ordering, by the applied magnetic field in the present electronic model \cite{com1}. Thus, the tendency of the PPB-induced AFM ordering is weaker compared with that in the last subsection. 

In this field configuration, both of the two coefficients $K_{\Delta \, m, \, n}$ ($n=1$ and $2$) in the coupling term change their sign with increasing the field and induce the PPB-induced AFM ordering. A typical phase diagram in this case is shown in Fig.4. The main feature such that the AFM ordering tends to be promoted by PPB is qualitatively the same as in ${\bf H} \perp {\hat {\bf n}}$. Reflecting the field-induced suppression of AFM order in the {\it normal} state mentioned above, however, the AFM ordering in this case is quantitatively weaker than that in ${\bf H} \perp {\hat {\bf n}}$, and the field region in which the AFM ordering is the most favorable tends to be shifted to lower fields than $H_{c2}(0)$. Our assumption in sec.I that, in CeCoIn$_5$, ${\hat {\bf n}}$ is locked to the $c$-axis irrespective of the direction of the field is closely related to this fact, because the AFM order just below $H_{c2}(0)$ is not realized in CeCoIn$_5$ in ${\bf H} \parallel c$ \cite{Kumagai06}. This issue will be discussed further in relation to Fig.9 again. 

\section{IV. Case with First order $H_{c2}$-Transition and AFM ${\bf Q}$-Vector}
In this section, we explain how the results in the last section are changed when the PPB is much stronger so that the $H_{c2}$-transition at lower temperatures is of first order. For brevity, the resulting SC state is assumed through this section to be spatially uniform at least in the direction parallel to ${\bf H}$. Consequently, the only additional transition in the SC state in high fields is the PPB-induced AFM ordering. We stress here that, as explained in sec.I, the resulting coexistent phase of the AFM order and the uniform $d$-wave SC one must not be identified with the HFLT phase of CeCoIn$_5$. The main purpose of this section is to clarify further details of the PPB-induced AFM ordering in addition to the obtained results in the last section. Relevance to the HFLT phase of CeCoIn$_5$ will be discussed in details in the next section. 

\begin{figure}[t]
\scalebox{0.5}[0.5]{\includegraphics{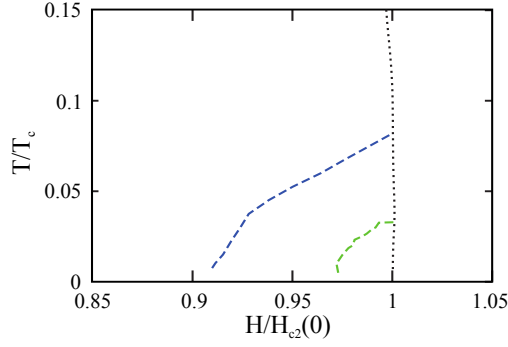}}
\caption{Upperlimits, defined by the vanishing of $\chi_s$, of possible AFM transition curves in ${\bf H} \perp c$ obtained in the perturbative approach for $\delta_{\rm IC}=0.44$ (lower curve) and $0.63$ (higher one). For each case, an actual transition curve is lower than the dashed curve depending on the value of the repulsive interaction $U$. The parameter values, $\gamma=4.5$ and $\alpha_{{\rm P}, \perp}=1.1$, are common between the two dashed 
curves.}
\end{figure}

\begin{figure}[t]
\scalebox{0.6}[0.6]{\includegraphics{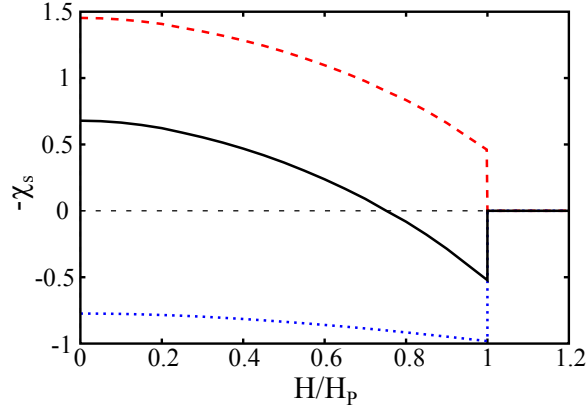}}
\caption{Typical $-\chi_s$ v.s. $H/H_{\rm P}$ curve (solid black curve) taken at $t=0.1$ in the strong PPB case in ${\bf H} \perp c$. This figure should be compared with Fig.2 with a second order $H_{c2}$-transition. The upper and lower dotted curves express the SC part of $-\chi^{(n)}$ and $-\chi^{(an)}$, respectively. Calculation was performed in the Pauli limit fully taking account of the $|\Delta|$ dependences and assuming the two-dimensional circular Fermi surface and the incommensurability $\delta_{\rm IC}=0.1$. }
\end{figure}

\subsection{${\bf H} \perp c$}

In most part of this section, we focus on the ${\bf H} \perp c$ case. 
First, let us start with pointing out an important difference between the PPB-induced AFM order and the ordinary itinerant AFM one in the normal state. As demonstrated in Fig.3, the AFM order usually diminishes with increasing pressure. However, applying the external pressure also enhances the incommensurability of the Fermi surface measured by $|\delta_{\rm IC}|$ or $|t_2|$. In relation to this, we present in Fig.5 two lines implying the $\delta_{\rm IC}$-dependence of the positions on which the SC part of the bare susceptibility $\chi_s \equiv \chi^{(n)} - \chi^{(n)}(\Delta=0) + \chi^{(an)}$ vanishes. This figure has been obtained by using a larger Maki parameter $\alpha_M \simeq 7$ leading to the first order $H_{c2}$-transition at low temperatures compared with that in Fig.3. Strictly speaking, a possibility of appearance of a FFLO state needs to be considered in the case of Fig.5. The situation with a coexistence of the AFM and FFLO orders will be discussed separately in the next section. 

Note that Fig. 5 shows that the present PPB-induced AFM ordering is enhanced with {\it increasing} the incommensurability. Since the normal part of the susceptibility, determining $T_{\rm N}$ used in Fig.5 as the key parameter dependent on the pressure, is not incorporated in Fig.5, and the pressure dependences of $T_{\rm N}$ and the incommensurability are competitive with each other for the PPB-induced AFM ordering, this figure means that, in the case with a stronger PPB, it cannot be concluded generally whether the AFM ordering occurring close to $H_{c2}(0)$ is enhanced or diminished with increasing pressure. 
\begin{figure}[t]
\scalebox{0.8}[0.8]{\includegraphics{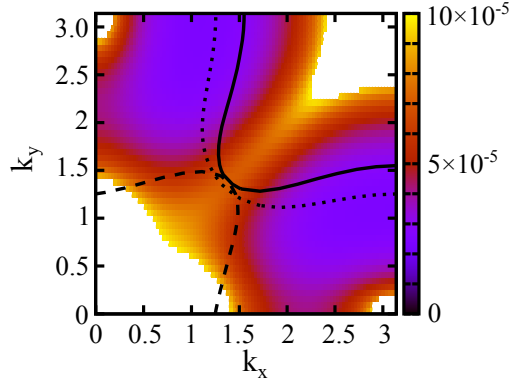}}
\caption{Fermi surface (solid curve) following from the dispersion relation, eq.(\ref{dispersionYamada}), with the values $t_1=20 T_c$, $t_2/t_1=-1.25$, $t_3/t_1=0.65$, and $\mu/t_1=1.85$ and leading to a diagonal wave vector of the incommensurate AFM order (see Fig.8 below) consistent with the experimental observation \cite{Kenzel1,Kenzel2}. The lower left dashed curve is the branch obtained by performing a ${\bf Q}_0 +$ ($0.37$, $0.37$) - shift for the original Fermi surface, while the dotted curve is the Fermi surface obtained \cite{HIkeda} by mimicing the result from the band calculation. For the ${\bf Q}_0 -$ ($0.37$, $0.37$)-shift, essentially the same nesting condition in the diagonal direction is obtained in the region $- \pi \leq k_x$, $k_y \leq 0$. Note that the magnitude of the density of states at each ${\bf k}=(k_x$, $k_y)$ is represented by the colors. }
\end{figure}

\begin{figure}[b]
\scalebox{0.6}[0.6]{\includegraphics{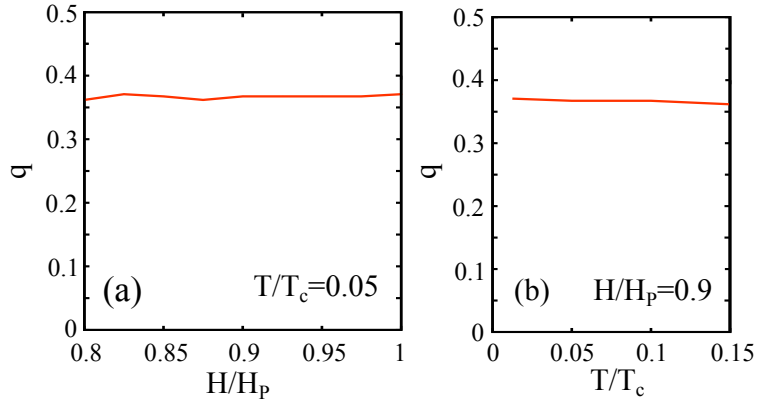}}
\caption{$H$ and $T$ dependences of the diagonal ${\bf q}$ resulting from the Fermi surface (solid curve) in Fig.7. This fact that ${\bf q}$ is nearly independent of $H$ and $T$ is consistent with the observation in Ref.\cite{Kenzel1,Kenzel2}, supporting the picture that the basic origin of making the incommensurate AFM wave vector in the HFLT phase of CeCoIn$_5$ diagonal consists in the electronic structures and the details of the Fermi surface. The corresponding phase diagram is given later in Fig.10 (b).}
\end{figure}

The corresponding data to Fig.2 in the last section are presented in Fig.6, where, for brevity, the orbital pair-breaking effect (the presence of the vortices) has been neglected. Reflecting the discontinuous nature of the $H_{c2}$-transition, $\chi_s$ discontinuously vanishes at $H_{c2}$ with increasing field. In the region in which $\chi_s > 0$, the PPB-induced AFM ordering is possible depending on the value of the normal part of the susceptibility. This figure clearly shows that the PPB-induced AFM ordering is not an artifact of the logarithmic divergence of the anomalous part upon cooling in the perturbative approach. 

Next, as one aspect representing the resulting AFM order, the incommensurate part ${\bf q}$ of the AFM modulation wave vector ${\bf Q} = {\bf Q}_0 + {\bf q}$ will be investigated. The optimal ${\bf q}$ is numerically found by replacing ${\bf Q}_0$ in eq.(\ref{fm2pauli}) and the ensuing expressions with ${\bf Q}_0 +{\bf q}$ and minimizing $f_m^{(2)}$ with respect to ${\bf q}$. Through our examination of possible ${\bf q}$, we have found that a diagonal ${\bf q}$ consistent with that determined experimentally in CeCoIn$_5$ \cite{Kenzel2} is not easily obtained in the case where the nesting condition on the Fermi surface is relatively kept even if deviating from the diagonal (i.e., ${\bf Q}_0$) direction. The dispersion relation (\ref{dispersionYamada}) with vanishing $t_3$ corresponds to this case. As shown in Fig.7, the Fermi surface with a remarkable inflection close to the diagonal direction is needed to obtain a diagonal ${\bf q}$ \cite{Kenzel2}. Here, we have assumed that the relevant Fermi surface to the $d$-wave superconductivity and the itinerant antiferromagnetism is the so-called $\gamma$-sheet \cite{HIkeda} or the band-14 electron one \cite{Onuki} which has the largest density of states and is {\it not} cylindrical. As examined previously \cite{RI072}, such a noncylindrical and 3D-like modulation of the Fermi surface is necessary to theoretically explain the presence of a longitudinal FFLO state in CeCoIn$_5$ in ${\bf H} \parallel c$ suggested from a NMR experiment \cite{Kumagai06}. On the other hand, it should be noted that, in contrast to the above-mentioned severe condition on the required modulation wave vector of the high field AFM order of CeCoIn$_5$, the AFM ordering itself close to $H_{c2}(0)$ is realized as a consequence of the strong PPB and the $d_{x^2-y^2}$-pairing symmetry irrespective of the details of the Fermi surface. 

In addition, we have also examined the field and temperature dependences of ${\bf q}$. As Fig.8 shows, the obtained diagonal ${\bf q}$ is mostly robust on sweeping the temperature and the magnetic field. This fact consistent with the experimental data \cite{Kenzel2} supports our picture that, in contrast to that in other work \cite{Yanase}, the origin of the AFM order in the HFLT phase of CeCoIn$_5$ is not a FFLO structure but of a purely electronic origin \cite{com2}. 

\begin{figure}[t]
\scalebox{0.6}[0.6]{\includegraphics{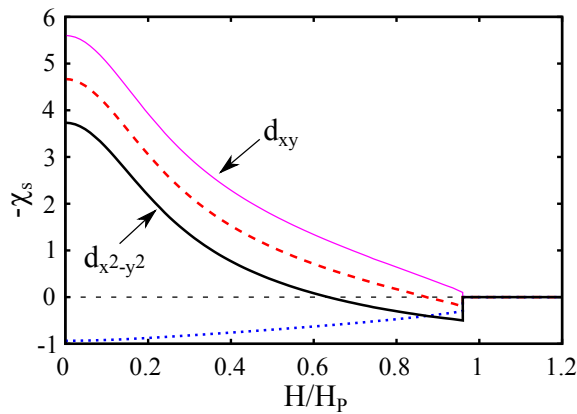}}
\caption{Curves including a $-\chi_s$ v.s. $H/H_{\rm P}$ curve (black solid curve) in ${\bf H} \parallel c$ which correspond to those in Fig.6. For comparison, the corresponding $-\chi_s(H)$ in $d_{xy}$-pairing case (uppermost curve) is also shown. Calculation was performed in the same scheme as that of Fig.6 but in the commensurate ($\delta_{\rm IC} \to 0$) limit \cite{com4}. }
\end{figure}

\subsection{${\bf H} \parallel c$}

 In obtaining Figs.6, 7, and 8, we have used the approach in the Pauli limit (see sec.II). As mentioned earlier, this approach is not necessarily unrealistic in examining thermodynamic data of quasi 2D materials in ${\bf H} \perp c$. In contrast, the corresponding neglect of the orbital pair-breaking in ${\bf H} \parallel c$ is usually unacceptable because of the important roles of the vortices in this configuration. Nevertheless, it will be useful to know the PPB-induced AFM ordering in the Pauli limit in this configuration. In Fig.9, we show an example of our results in ${\bf H} \parallel c$ obtained in the Pauli limit. In contrast to Fig.4, the field at which the AFM ordering is the most remarkable is much closer to $H_{c2}(0)$, irrespective of the position of an AFM-QCP suggested from the data in the normal state (see Fig.1 and the discussion relevant to Fig.4 in sec.III). We believe that this difference between Fig.4 and 10 is intrinsic and is a reflection of the difference in the magnitude of PPB. 
To clarify this issue further, one would need to perform a more elaborate analysis taking account of both the PPB and the orbital pair-breaking on an equal 
footing in future. 

In Fig.9, the corresponding $-\chi_s(H)$ curve in the case of $d_{xy}$-pairing has also been presented, for comparison, which implies that, even in ${\bf H} \parallel c$, the $d_{xy}$-pairing case does not lead to the PPB-induced AFM ordering. This close relation between the diagonal AFM ${\bf Q}$-vector and the direction of the gap node indicates that the four-fold symmetric $d$-wave SC symmetry in a SC material showing an AFM ordering enhanced on approaching $H_{c2}(0)$ from below with increasing field should be always the $d_{x^2-y^2}$-one.  

\section{V. Effect of FFLO modulation on AFM Ordering in Pauli Limit} 

In the preceding sections, we have focused on the case with a finite AFM transition temperature in some field range just below $H_{c2}(0)$. From such a case, the situation with a remarkable AFM critical fluctuation near $H_{c2}(0)$ but with no genuine AFM order is easily created by reducing the repulsive interaction strength $U$. As mentioned in Introduction, however, when discussing the HFLT phase of CeCoIn$_5$ in ${\bf H} \perp c$ with ${\hat {\bf n}} \parallel c$, it is indispensable to, consistently, take account of appearance there of a different ordered state from the AFM order induced by the strong PPB, such as the spatial modulation of the SC order parameter of the longitudinal FFLO state. 

Here, it is pointed out that the FFLO spatial modulation of the SC order parameter in ${\bf H} \perp c$ significantly enhances the PPB-induced AFM ordering. In the context of CeCoIn$_5$, it is an elaborate task to give a full description of this interplay of the two orderings within the two approaches explained in sec.II. Here, we examine this interplay by assuming the wavelength of the FFLO modulation $2 \pi/q_{\rm LO}$, defined through eq.(\ref{LOdelta}), to be long enough to neglect the gradient terms in the free energy on the AFM ordering. Implication of this local approximation on the AFM modulation will be explained at the end of this section. 

\begin{figure}[t]
\scalebox{1.0}[1.0]{\includegraphics{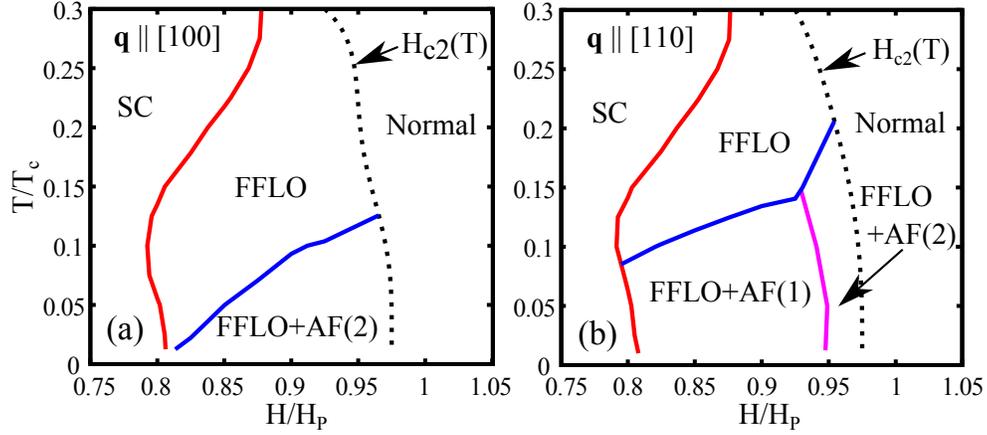}}
\caption{Theoretical $H$-$T$ phase diagrams in ${\bf H} \perp c$ following from our calculation in the Pauli limit. The symbols SC, FFLO, and AF ($n$) denote the uniform SC order (see the text), the FFLO one, and the AFM one defined in Fig.11, respectively. The $H_{c2}$-transition on the dotted curve is discontinuous, while the FFLO and AFM orders disappears continuously on the red solid and blue solid curves, respectively. The used values of the parameters are $t_1=\mu=20 T_c$, $t_2/t_1=-1.25$, and $t_3/t_1=0.35$ in (a), while, in (b), they are the same as those in Fig.7. 
In the direction parallel to ${\bf H}$, the AFM order in (a) has the in-phase structure relative to the FFLO modulation in any field range (see Fig.11 (2) below). However, this AFM order is lost (i.e., $m$ vanishes) in the FFLO state with finite $q_{\rm LO}$, and the incommensurate part, ${\bf q}$, of the AFM wavevector ${\bf Q}$ is parallel to [1,0,0], in contrast to the observation \cite{Kenzel1,Kenzel2}. On the other hand, in (b), the AFM order shows a structural transition \cite{com3} between the two structures shown in Fig.11 within the FFLO state, and, with decreasing $H$, the AFM order is continuously lost on the red solid curve in the manner accompanied by the disappearance of the FFLO nodal planes. The corresponding ${\bf q}$ is shown in Fig.8 and is parallel to [1,1,0] as seen in experiments \cite{Kenzel1,Kenzel2}.}
\end{figure}

\begin{figure}[b]
\scalebox{0.8}[0.8]{\includegraphics{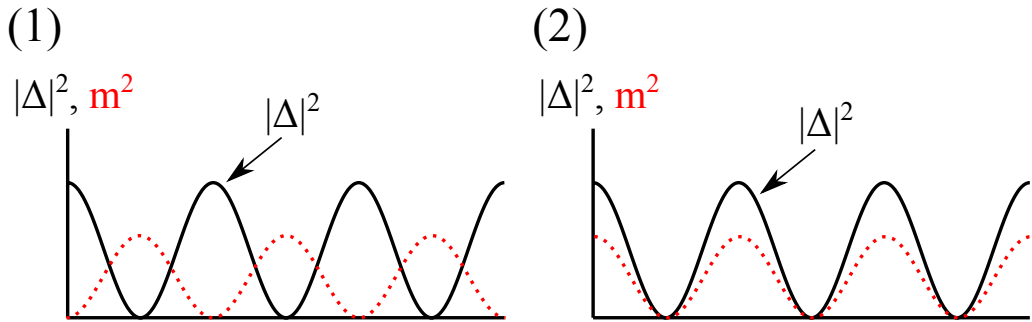}}
\caption{Two structures, (1) and (2), of the spatial modulation of the AFM order parameter $m$ relative to that of $\Delta$ in the coupled FFLO and AFM phase. The AF($n$) ($n=1$ and $2$) in Fig.10 corresponds to the structure ($n$) in this figures. For simplicity, we have described the AFM order in the form ${\rm sin}(q_{\rm LO}x + \phi_0)$. The structure (1) ((2)) corresponds to the case with $\phi_0=0$ ($\phi_0=\pi/2$). The phase $\phi_0$ continuously changes with increasing $H$ in the interval between zero and $\pi/2$. }
\end{figure}

Possible phase diagrams including the FFLO and AFM ordered states and resulting from our calculation in the Pauli limit are shown in Figs.10 and 12. There, the uniform SC state in lower fields in the Pauli limit corresponds to the ordinary vortex lattice in the full description including the orbital pair-breaking effect. In the present formulation in the Pauli limit, a possible FFLO state is found by substituting the test solution eq.(\ref{LOdelta}) of the SC order parameter $\Delta$ into the free energy terms given by eq.(\ref{FSCgrad}) and minimizing them with respect to $q_{\rm LO}$. Then, in our formulation in the Pauli limit, the $H_{c2}$-transition was of first order, while the resulting transition between the FFLO and uniform SC states was of second order at any temperature, implying that the distance between the neighboring FFLO nodal planes diverges at the transition. Note that, in considering the FFLO ordering, the averaged value of $|\Delta|$ is so rigid that it may be assumed to be unaffected by an AFM ordering. 

\begin{figure}[t]
\scalebox{0.9}[0.9]{\includegraphics{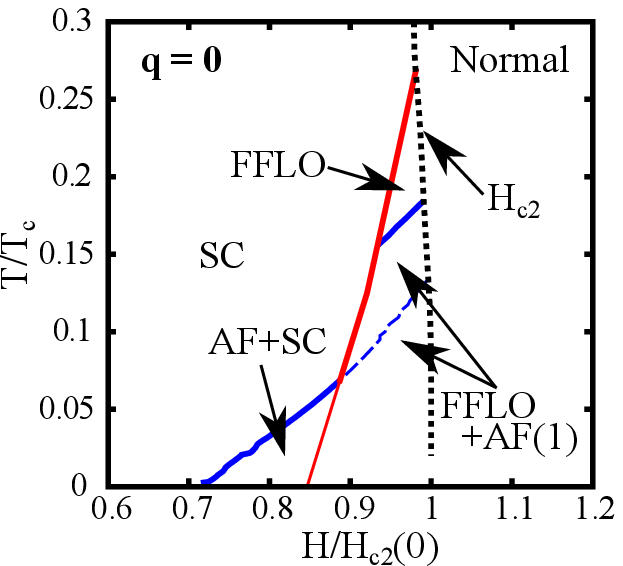}}
\caption{Another theoretically possible phase diagram including the FFLO and AFM orders following from the parameter values, $t_1=10 T_c$, $t_2/t_1=0.02$, and $t_3= \mu=0$. The blue solid curves are the actual AFM transition lines, and, in the absence of the FFLO state, the dashed blue becomes the AFM transition line at higher fields. This case in which the AFM order appears without the FFLO state in lower fields does not apply to the high field phase diagram of CeCoIn$_5$ (see the text for details). }
\end{figure}

Then, when considering a coexistence of the AFM and FFLO orderings in the local approximation for the AFM modulation, we only have to use eqs.(\ref{fm2perpn}) and (\ref{fm4perpn}) with $\Delta$ replaced with $\Delta({\bf r})$ given in eq.(\ref{LOdelta}). We have checked that, in all cases we have examined, $f_m^{(4)} > 0$ so that the AFM transition is continuous by itself. First, it will be clarified how spatial variation of $m$ should be realized in the LO structure, eq.(\ref{LOdelta}), of $\Delta$. It is easily understood by recalling the role of the coupling term $f_{\Delta \, m}^{(2,2)}$ in the perturbative approach that, when $f_{\Delta \, m}^{(2,2)} > 0$ ($< 0$), the structure of the AFM order parameter showing the our-of-phase (in-phase) modulation for the SC order parameter is the most stable, and that the spatially uniform AFM order is unstable in the FFLO state. It is directly concluded from this consideration that, in the presence of the Larkin-Ovchinnikov state, eq.(\ref{LOdelta}), the uniform AFM order cannot be stable, and that, at least in higher fields where $|\Delta|$ is smaller, the in-phase structure, Fig.11 (2), is the most stable as a direct consequence of the PPB-induced AFM ordering. On the other hand, in lower fields and particularly close to the second order transition to the uniform SC state, we have a couple of candidates of possible phase diagrams, and it is not easy to predict which of them should be realized in a particular material. In fact, to clarify the best candidate for CeCoIn$_5$, extensive consideration is necessary as follows. First, an experimental fact \cite{Tokiwacom} that the anomalous doping (impurity) effect on the second order transition in CeCoIn$_5$ \cite{Tokiwa1,Tokiwa2}, indicating the presence of the FFLO state above the transition \cite{RI10}, is seen entirely over the field range occupied by the HFLT phase implies that CeCoIn$_5$ at ambient pressure does not show realization of the phase diagram of the type of Fig.12 in which a direct transition between the AFM and the uniform SC phases occurs {\it without} the FFLO state at lower temperatures. On the other hand, it is not easy to theoretically justify a simultaneous disappearance of the in-phase AFM order, sketched in Fig.11 (2), and the FFLO one at the same {\it second order} transition where $2 \pi/q_{\rm LO}$ diverges: Within this scenario, a simultaneous disappearance of the two orders would require a discontinuous vanishing of the AFM order parameter $|m|$ in contrast to the observations in CeCoIn$_5$. For this reason, we propose two candidates, following from our microscopic calculations, of the high field phase diagram of CeCoIn$_5$ in Fig.10. In (a), the PPB-induced in-phase AFM order, sketched in Fig.11 (2), is lost within the FFLO state just above the second order transition, while the longitudinal modulation of the AFM order changes from the in-phase modulation to the out-of-phase one sketched in Fig.11 (1) on approaching the second order transition, as a result of the reduction of PPB. The out-of-phase AFM modulation is consistent with the conventional picture \cite{Yanase,RIprlcom} that a spatial region where $|\Delta|$ is small is occupied by a competitive non SC order. In this case, the AFM order rides on the nodal planes and thus, is continuously lost through the continuous disappearance of the FFLO nodal planes, while the PPB mechanism of the AFM order favors the {\it local} coexistence with the SC order, i.e., a structure like Fig.11 (2). In any case, an additional continuous transition \cite{com3} inevitably appears within the resulting HFLT phase of the theoretical phase diagrams appropriate for CeCoIn$_5$. If taking account of the consistency on the direction of AFM-${\bf Q}$ vector between the experiment \cite{Kenzel1,Kenzel2} and our results, Fig.10 (b) with the diagonal ${\bf q}$ shown in Fig.8, becomes the best candidate as the phase diagram of CeCoIn$_5$, while the electronic parameters (see the caption of Fig.10) resulting in Fig.10 (a) never leads to a ${\bf q}$ in the diagonal direction. However, this correspondence between the longitudinal structure of the AFM order and the detailed direction of the AFM ${\bf Q}$-vector is quite a subtle issue and might be changed due to a further refinement of the starting electronic model. For this reason, it appears that we should not conclude here which of the two figures in Fig.10 is more appropriate for CeCoIn$_5$. 

The presence of the wide FFLO phase with {\it no} AFM order at higher temperatures in Fig.10 might indicate that the present picture on the HFLT phase of CeCoIn$_5$ is insufficient, because such a nonmagnetic FFLO region has not been identified so far in experiments on this material. However, the present analysis in the Pauli limit where $\alpha_{{\rm P}, \perp} = \infty$ certainly overestimated the temperature region of the FFLO phase. In an improved analysis for finite $\alpha_{{\rm P}, \perp}$-values to be performed in future works, the FFLO region unaccompanied by the AFM order is certainly expected to be narrower if taking account of the orbital pair-breaking within the present approach. 

Here, we comment on justification of our use of the local approximation on the AFM modulation. As far as the transition between the FFLO and uniform SC states is of second order, this local approximation is always 
justified at least just above the transition because the order parameter \cite{ada} of this transition, $q_{\rm LO}$, is inversely proportional to the distance between the neighboring nodal planes which diverges at the transition. At higher fields, however, $q_{\rm LO}$ grows so that the neglect of the gradient terms may not be justified in general. If so, the above results in the local approximation would be quantitatively changed by the gradient terms. In the case of CeCoIn$_5$, however, the presence of the strong PPB-induced AFM fluctuation induces \cite{RI072} a quasiparticle damping which destabilizes the transverse FFLO states to be described by {\it higher} Landau level modes of the SC order parameter compared with the longitudinal FFLO state, eq.(\ref{LOdelta}). Further, this increase of the quasiparticle damping even weakens effects of PPB and results in a reduction of $q_{\rm LO}$. We expect the use of the local approximation for the AFM ordering to be justified in this sense. 

It should also be noted that the spatial variation due to the vortices in the real systems with the orbital pair-breaking has not been incorporated in this section. In fact, as in the AFM ordering detected in the vortex core of high $T_c$ cuprates \cite{Kumagaihightc}, it is natural to expect appearance of an AFM order in the spatial region where the SC order is weaker like the vortex core. Even in the present case with strong PPB, the spatial modulation perpendicular to ${\bf H}$ of the SC order parameter due to the vortex structure assists the AFM ordering, although its effect is found to be weaker than that of the FFLO modulation. On the other hand, contrary to the event seen in cuprates \cite{Kumagaihightc}, the present PPB-induced AFM order tends to coexist with the SC order so that the AFM order parameter value $|m|$ is maximal {\it outside} the vortex core. Details of this issue will be reported elsewhere \cite{aoyama}. 

\section{VI. Summary and Concluding Remarks}

The phase diagrams shown in Fig.10 are comparable with that following from a recent experiment \cite{Kumagai11}. In the NMR measurement in Ref.\cite{Kumagai11}, the presence of {\it both} the AFM order and the normal state region has been detected in the HFLT phase of CeCoIn$_5$ at $50$ (mK) in ${\bf H} \perp c$, and the square-root field dependence of the quasiparticle number consistent with that of the FFLO order parameter, i.e., $q_{\rm LO} \propto \sqrt{H-H_2(T)}$, has been found, where $H_2(T)$ is the field at which the second order transition to the ordinary vortex lattice phase (the uniform SC phase in the present Fig.10) occurs. Another crucial observation in Ref.\cite{Kumagai11} is that, in contrast to the picture in Ref.\cite{Yanase}, the observed AFM order is extended spatially without being localized in a narrow spatial region. It appears that the in-phase structure in Fig.11 (2), i.e., the coexistence of the AFM and SC orders induced by PPB, is the consistent picture with this experimental fact. 

In sec.IV, we have examined the direction of the modulation wave vector ${\bf Q}$ of the stable AFM order originating from the PPB-induced mechanism and have found that, including its incommensurate component ${\bf q}$, ${\bf Q}$ can take the orientation parallel to the direction of the gap node irrespective of the ${\bf H}$-direction in the $a$-$b$ plane. To the best of our knowledge, this is the first study giving a consistent calculation result on ${\bf q}$ with the data \cite{Kenzel1,Kenzel2} in the HFLT phase of CeCoIn$_5$. In relation to this, we note that we could not find a diagonal ${\bf q}$-vector in terms of the more familiar tight-binding electronic Hamiltonian \cite{Aperis} with no $t_3$ term in the range of the values we have assumed for the parameters $t_1$, $t_2$, and $\mu$. In any case, it should be stressed that this ${\bf Q}$-direction is highly sensitive to the details of the starting electronic model. 

Recent neutron scattering experiments \cite{Eskildsen10} on CeCoIn$_5$ in the field directions tilted from the $a$-$b$ plane have shown that just 17 degrees' tilt of the applied field results in disappearance of the AFM order present in ${\bf H} \perp c$. Note that the focus of this experiment is the AFM order and not the tilt direction signaling disappearance of the HFLT phase, i.e., of the FFLO order. It is interesting to note that, according to another observation \cite{Correa} resulting from the field tilt, the HFLT phase survives up to 20 degrees, suggestive of the presence of the FFLO state with no AFM order in a narrow high angle region. In relation to this, we point out that such an AFM order disappearing separately from the FFLO order due to the field tilt is not surprising from the view point of the present theory, because, as is seen by comparing Fig.3 with Fig.4, the PPB-induced AFM ordering in ${\bf H} \parallel c$ is much weaker than that in ${\bf H} \perp c$ as far as, as have been assumed throughout this paper, ${\hat {\bf n}}$ is locked to the $c$-axis. A further theoretical study on this issue may be useful for confirming the genuine picture on the HFLT phase. 

In relation to this disappearance of the AFM order due to the field-tilt, a mechanism of AFM ordering of CeCoIn$_5$ below $H_{c2}$ has been argued in Ref.\cite{Suzuki} where it results from the four-fold symmetric enhancement of the density of states in the vortex lattice. However, the Fermi surfaces assumed there \cite{Suzuki} to support the AFM order are the nearly cylindrical band 15-electron ones, in the notation of Ref.\cite{Onuki}, with a smaller density of states, while the origin of the $d$-wave superconductivity and the FFLO state in ${\bf H} \parallel c$ \cite{Kumagai06} is the noncylindrical (3D-like) \cite{RI072} band 14-electron Fermi surface \cite{Onuki} corresponding to that of Fig.7 (see the text of sec.IV). It is unreasonable for the Fermi surface relevant to superconductivity to change with tilting the magnetic field. While preparing the final version of our manuscript, we were aware of another proposal \cite{Kato} on the AFM ordering in ${\bf H} \perp c$ below $H_{c2}$ in which attention is paid to the Zeeman-splitted nodal quasiparticles as the origin of the AFM order. Judging from the similarity on the starting model, the contribution mentioned in Ref.\cite{Kato} to the AFM ordering should be already included in the present theory with no limitation on the quasiparticles close to the gap nodes. However, we remark that focusing \cite{Kato} on the Zeeman-splitted nodal quasiparticles would result in a remarkable field dependence of the incommensurate component ${\bf q}$ of the AFM wave vector ${\bf Q}$, in contrast to our result in Fig.8 consistent with the observation \cite{Kenzel2}. Further, we stress here that the argument in Ref.\cite{RI10}, given in relation to Fig.2 there, is also applicable to any picture \cite{Suzuki,Kato} identifying the experimental second order transition with a pure AFM transition and thus that the doping effect \cite{Tokiwa1,Tokiwa2} leading to an extremely dramatic suppression of the transition with {\it no} notable change of the transition point to the HFLT phase is incompatible with such purely AFM scenarios on the second order transition (see also sec.I). 

In this manuscript, we have also stressed a close relation between the ${\bf Q}$-vector of the AFM order or fluctuation enhanced close to $H_{c2}(0)$ and the nodal direction of the $d$-wave SC energy gap. Thus, if a novel SC material is accompanied by such an AFM ordering in high fields, it may become a useful method for obtaining information on the nodal direction of the SC pairing symmetry. We wish to stress that, in contrast to measurements of the thermal conductivity \cite{Izawa} and the specific heat \cite{Sakaki} in which experiments at very low temperatures have been necessary to determine the pairing symmetry, the present method does not require measurements at such low enough temperatures to obtain knowledge on the pairing symmetry. 

\section{Acknowledgement} 

We thank K. Aoyama, K. Kumagai, Y. Matsuda, R. Movshovich, H. Shishido, T. Shibauchi, M. Sigrist, and Y. Yanase for useful discussions. 
This work was partly supported by the Grant-in-Aid for Scientific Research [No. 21540360] from JSPS, Japan.

\section{Appendix} 

Here, detailed expressions on $f_m^{(4)}$ and $f_{\Delta m}^{(2,4)}$ to be used in determining the character of the AFM transition in the perturbative approach will be listed. As far as the orbital pair-breaking effect is neglected in $f_{\Delta m}^{(2,4)}$, they are expressed by 
\begin{widetext}
\begin{equation} f_{m}^{(4)} = \frac{1}{2 \beta} \sum_{\varepsilon_n, {\bf k},\sigma} \biggl[ \mathcal{G}^{(\sigma)}_{\varepsilon_n}({\bf k}) \mathcal{G}^{(\bar{\sigma})}_{\varepsilon_n}({\bf k}+{\bf Q}_0) \mathcal{G}^{(\sigma)}_{\varepsilon_n}({\bf k}) \mathcal{G}^{(\bar{\sigma})}_{\varepsilon_n}({\bf k}+{\bf Q}_0) \biggr] m^4, 
\label{eq:Fm4}
\end{equation}
and 
\begin{eqnarray} 
f_{\Delta m}^{(2,4)} &=& \beta^{-1} \sum_{\varepsilon_n, {\bf k},\sigma} |w_{\bf k}|^2 \biggl[ \, [ \mathcal{G}^{(\sigma)}_{\varepsilon_n}({\bf k}) ]^3 \, [ \mathcal{G}^{(\bar{\sigma})}_{\varepsilon_n}({\bf k}+{\bf Q}_0) ]^2 \, \mathcal{G}^{(-\sigma)}_{-\varepsilon_n}(-{\bf k}) + \frac{\sigma \bar{\sigma}}{2} \, [ \mathcal{G}^{(\sigma)}_{\varepsilon_n}({\bf k}) ]^2 \, [ \mathcal{G}^{(\bar{\sigma})}_{\varepsilon_n}({\bf k}+{\bf Q}_0) ]^2 \nonumber \\
&\times& \mathcal{G}^{(-\bar{\sigma})}_{-\varepsilon_n}(-{\bf k}+{\bf Q}_0) \, \mathcal{G}^{(-\sigma)}_{-\varepsilon_n}(-{\bf k}) \nonumber \\
&+& [ \mathcal{G}^{(\sigma)}_{\varepsilon_n}({\bf k}) ]^2 \, [ \mathcal{G}^{(-\sigma)}_{-\varepsilon_n}(-{\bf k})]^2 \, \mathcal{G}^{(\bar{\sigma})}_{\varepsilon_n}({\bf k}+{\bf Q}_0) 
\, \mathcal{G}^{(-\bar{\sigma})}_{-\varepsilon_n}(- {\bf k}+{\bf Q}_0) \biggr] \langle |\Delta|^2 m^4 \rangle_{sp}.  
\label{eq:FD2m4}
\end{eqnarray}
\end{widetext}

Their expressions to be useful in numerical analysis are in the following. 
In ${\hat {\bf n}} \parallel {\bf H}$, they are 
\begin{equation} 
f_{m}^{(4)} = - \frac{N(0) \beta}{32 \pi^2} \sum_{\sigma} \mathrm{Re} \biggl[ \psi^{(2)}\biggl( \frac{1}{2}+i\frac{I\sigma \beta}{2 \pi}+i\frac{\delta \beta}{4 \pi} \biggr) \biggr] m^4,  
\label{eq:m4para}
\end{equation}
and 
\begin{widetext}
\begin{eqnarray}
f_{\Delta m}^{(2,4)} &=& N(0) \sum_{\sigma} \biggl\{ 
\frac{1}{16 \delta} \biggl(\frac{\beta}{2\pi} \biggr)^3 \mathrm{Im} \biggl[ \psi^{(3)}\biggl( \frac{1}{2}+i\frac{I \beta \sigma}{2 \pi}-i\frac{\delta \beta}{4 \pi} \biggr) \biggr] - \frac{3}{8\delta^2} \biggl(\frac{\beta}{2 \pi} \biggr)^2 \mathrm{Re} \biggl[ \psi^{(2)}\biggl( \frac{1}{2}+i\frac{I \beta \sigma}{2 \pi} -i\frac{\delta \beta}{4 \pi} \biggr) \biggr] \nonumber \\ 
&-& \frac{3 \beta}{4 \pi \delta^3} \mathrm{Im} \biggl[ \psi^{(1)} \biggl( \frac{1}{2}+i\frac{I \beta \sigma}{2 \pi}-i\frac{\delta \beta}{4 \pi} \biggr) \biggr] + \frac{3}{\delta^4} \mathrm{Re} \biggl[ \psi\biggl( \frac{1}{2}+i\frac{I \beta \sigma}{2 \pi}-i\frac{\delta \beta}{4 \pi} \biggr) - \psi\biggl( \frac{1}{2}+i\frac{I \beta \sigma}{2 \pi} \biggr) \biggr] \biggr\} |\Delta|^2 m^4, 
\label{eq:D2m4para}
\end{eqnarray}
\end{widetext}
while, in ${\hat {\bf n}} \perp {\bf H}$, they become 

\begin{equation} 
f_{m}^{(4)} = - N(0) \biggl(\frac{\beta}{4 \pi} \biggr)^2 \mathrm{Re} \biggl[ \psi^{(2)}\biggl( \frac{1}{2}+i\frac{\delta \beta}{4 \pi} \biggr) \biggr] m^4, 
\label{eq:m4perp}
\end{equation}
and 
\begin{widetext}
\begin{eqnarray}
f_{\Delta m}^{(2,4)} &=& N(0) \biggl\{ \frac{\delta}{8(\delta^2 - 4 I^2)} \biggl(\frac{\beta}{2 \pi} \biggr)^3 \mathrm{Im}\biggl[ \psi^{(3)}\biggl( \frac{1}{2}-i\frac{\delta \beta}{4 \pi} \biggr) \biggr] + \frac{2 I^2}{(4I^2-\delta^2)^2} \nonumber \\
&\times& \biggl(\frac{\beta}{2 \pi} \biggr)^2 \mathrm{Re}\biggl[ \psi^{(2)}\biggl(\frac{1}{2}+i\frac{I \beta}{2 \pi} \biggr) - \psi^{(2)}\biggl(\frac{1}{2}-i\frac{\delta \beta}{4 \pi} \biggr) \biggr] \biggr\} \langle |\Delta|^2 m^4 \rangle_{sp}.
\label{eq:D2m4perp}
\end{eqnarray}

\end{widetext}
Here, $\psi^{(n)}(z)$ is the $n$-th polygamma function satisfying 
\begin{equation} 
\psi^{(n)}(z) = (-1)^{n+1} n! \sum_{s=0}^\infty \frac{1}{(s+z)^{n+1}}.
\end{equation}


\begin{thebibliography}{99}
\bibitem{com3} Since just two structures are compared in Fig.11, one might feel that the structural transition on the AFM order should be discontinuous. However, we have verified by introducing a phase variable interpolating the two structures of Fig.11 for a single set of parameters in the electronic Hamiltonian that the transition occurs through a continuous change of the phase variable (see the caption of Fig.11). Its details will be reported elsewhere.
\end{thebibliography}
\end{document}